\documentclass[prl,twocolumn,aps,preprintnumbers,nofootinbib]{revtex4-2}
\usepackage{graphicx}
\usepackage{color}
\usepackage{bm}
\usepackage{amsmath}
\usepackage{amssymb}
\usepackage{xspace}
\usepackage[normalem]{ulem}
\usepackage{units}
\usepackage{dcolumn}
\usepackage{paralist}
\usepackage{natmove}
\usepackage{natbib}
\usepackage{mathptmx}
\DeclareSymbolFont{epsilon}{OML}{ntxmi}{m}{it}
\DeclareMathSymbol{\Epsilon}{\mathord}{epsilon}{"0F}

\usepackage{hyperref}

\begin{document}

\title{Effect of Electron-Phonon and Electron-Impurity Scattering on Electronic Transport Properties of Silicon/Germanium Superlattices}

\author{Manoj~Settipalli}
\affiliation{Ann and H.J. Smead Aerospace Engineering Sciences, University of Colorado Boulder, Boulder, CO 80303, USA}

\author{Vitaly~S~Proshchenko}
\affiliation{Ann and H.J. Smead Aerospace Engineering Sciences, University of Colorado Boulder, Boulder, CO 80303, USA}

\author{Sanghamitra Neogi}
\email{sanghamitra.neogi@colorado.edu}
\affiliation{Ann and H.J. Smead Aerospace Engineering Sciences, University of Colorado Boulder, Boulder, CO 80303, USA}


\begin{abstract}
Semiconductor superlattices have been extensively investigated for thermoelectric applications, to explore the effects of compositions, interface structures, and lattice strain environments on the reduction of thermal conductivity, and improvement of efficiency. Most studies assumed that electronic properties of superlattices remain unaffected compared to their bulk counterparts. However, recent studies demonstrated that electronic properties of silicon (Si)/germanium (Ge) superlattices show significant variations depending on compositions and growth substrates. These studies used a constant relaxation time approximation, and ignored the effects of electron scattering processes. Here, we consider electron scattering with phonons and ionized impurities, and report first-principles calculations of electronic transport properties including the scattering rates. We investigate two classes of Si/Ge short-period superlattices: superlattices with varied compositions grown on identical substrates and with identical compositions but grown on different substrates. We illustrate the relationship between the energy bands of the superlattices and the electron-phonon relaxation times. We model the electron-ionized impurity interaction potentials by explicitly accounting for the in-plane and the cross-plane structural anisotropy of the configurations. Our analysis reveals that the inclusion of electron-phonon and electron-impurity scattering processes can lead to $~\sim{1.56}$-fold improved peak power-factor of superlattices, compared to that of bulk Si. We observe that superlattices can also display dramatically reduced power-factors for certain lattice strain environments. Such reduction could cancel out potential thermoelectric efficiency improvements due to reduced thermal conductivities. Our study provides insight to predict variation of electronic properties due to changes in lattice strain environments, essential for designing superlattices with optimized electronic properties. 
\end{abstract}


\maketitle

\section{Introduction}
Semiconductor nanostructures, in particular silicon (Si)/germanium (Ge) based materials have risen to prominence over the past few decades owing to their applications in key technologies, including electronics~\cite{thompson200490,meyerson1994high,nissim2012heterostructures,paul2004si}, optoelectronics~\cite{koester2006germanium,liu2010ge}, thermoelectrics~\cite{alam2013review,taniguchi2020high,chen2003recent,dresselhaus2007new}, spintronics~\cite{awschalom2013quantum,jansen2012silicon}, and quantum applications~\cite{zwanenburg2013silicon}. Nanostructuring approaches have opened remarkable possibilities to design novel Si/Ge based materials, with tailored electronic and thermal properties, to enable these applications. A thermoelectric material operates at its peak efficiency when the thermoelectric figure of merit, ZT$=\text{S}^2 \sigma \text{T}/\kappa$, is maximized. Here, S is the Seebeck coefficient or thermopower, $\sigma$ is the electrical conductivity, $\kappa$ is the electronic and the lattice thermal conductivity, and T is the absolute temperature. Over the past two decades, several nanostructuring strategies have been proposed to reduce the thermal conductivity of Si based thermoelectric materials, and thus, improve ZT. Some examples include two-dimensional silicene~\cite{zhang2014thermal}, Si-Ge nanomeshes~\cite{perez2016ultra}, layered-graded Si$_{1-x}$Ge$_x$/Si superlattices (SL)~\cite{reparaz2012influence}, and Si membranes with and without surface nanostructures~\cite{neogi2015tuning,neogi2015thermal,mangold2016optimal,xiong2017native,neogi2020anisotropic,honarvar2018two,davis2014nanophononic}. Si/Ge based SLs, in particular, have been extensively studied to maximize the reduction of thermal conductivity by varying layer compositions~\cite{hahn2016effect,ferrando2015tailoring,zhang2017impeded}, interface structures~\cite{chen2013role,samvedi2009role,li2012effect}, and lattice strain environments~\cite{chen2013role,samvedi2009role,li2012effect}. Although most studies focused on identifying strategies to reduce $\kappa$, there has been an increasing interest in improving the electronic power-factor (PF), S$^2 \sigma$. The combination of reduced thermal conductivity and improved PF promises to greatly improve the thermoelectric efficiency of a nanostructured material. 

Prediction of electronic properties of materials, using first-principles density functional theory (DFT)-based approaches, has seen rapid progress in recent years. However, it still remains a challenge to predict electronic transport properties of nanostructured materials using these approaches. The structural variability introduced by non-uniform strain environments, dislocations and other growth dependent parameters is often too difficult to represent in first-principles models due to large computational expenses, making the prediction unreliable. As a consequence, only a few first-principles studies analyzed the electronic structure~\cite{satpathy1988electronic,hybertsen1987theory,ciraci1988strained,bernard1991strain} or transport properties~\cite{hinsche2012thermoelectric,shi2012high,settipalli2020theoretical,proshchenko2019optimization,proshchenko2021role,pimachev2021first} of even planar Si/Ge SLs. Electronic transport properties of nonpolar semiconductors are significantly affected by electron scattering due to lattice vibrations, ionized impurities, neutral impurities, dislocations, vacancies, and interstitials~\cite{dekker1981nonpolar,nag2012electron,gupta2020theoretical}. The effect of various scattering mechanisms on electronic and thermal properties of Si nanostructures has been acknowledged in previous studies~\cite{fu2017electron,yang2017thermal,xu2019effect}. However, most of these studies used carrier concentration-dependent electron relaxation times~\cite{liao2015significant,mangold2016optimal,hinsche2012thermoelectric,proshchenko2019heat}, parametrized by the experimental bulk Si mobility data~\cite{jacoboni1977review,baccarani1975electron}. The effects of electron-phonon scattering on thermal conductivities and the resultant ZT of bulk~\cite{liao2015significant} and nanocrystalline~\cite{yang2017thermal} Si, Si nanostructures~\cite{fu2017electron}, SiGe alloys~\cite{yang2017thermal,xu2019effect}, and SiGe compounds~\cite{fan2018first}, have been investigated with first-principles approaches. In comparison, the role of the different scattering processes on electronic transport properties of Si/Ge based materials have received little attention~\cite{murphy2008first}. In particular, the effects of different electron scattering processes on the electronic transport coefficients of Si/Ge based superlattices, such as the Seebeck coefficients, the electronic resistivities or the power-factors, are not established.

In this article, we illustrate the effects of electron-phonon (EPS) and electron-ionized impurity scattering (IMS) processes on the electronic transport properties of n-type [001]-Si/Ge based superlattices. We investigate two classes of eight-atom Si/Ge SLs with diverse lattice strain environments: SLs with varied layer compositions grown on same substrate and SLs with identical compositions but grown on different substrates. The role of lattice strain and interface roughness on electronic properties of p-type Ge/Si$_{0.5}$Ge$_{0.5}$~\cite{ferre2013cross, samarelli2013thermoelectric} and n-type SiGe/Si SLs~\cite{taniguchi2020high,koga2000experimental} has been highlighted by experimental studies. In our recent first-principles studies~\cite{settipalli2020theoretical,proshchenko2019optimization,proshchenko2021role,pimachev2021first}, we showed that the cross-plane electronic transport properties of n-type [001]-Si/Ge SLs can be tuned by varying the SL layer compositions, periods, and growth substrates. We noted that such nanostructuring strategies positively impact the PF in the high-doping regime. However, the past studies primarily employed the constant relaxation time approximation (CRTA) within the BTE framework. Here, we calculate the electronic structure properties of the Si/Ge SLs by performing DFT calculations, as implemented in the Quantum Espresso package~\cite{giannozzi2009quantum}. We compute the EPS and IMS rates using a perturbative approach that follows Fermi's golden rule. The electron-phonon matrix elements are computed using the electron-phonon Wannier (EPW) package based on maximally localized Wannier functions~\cite{giustino2007electron,ponce2016epw}. This approach provides accurate interpolation of the matrix elements from coarse grids to dense grids~\cite{park2014electron}, and thus reduces computational expenses. We calculate the cross-plane electronic transport properties (S, $\rho$, PF) of the Si/Ge SLs by incorporating the energy dependent electron scattering rates within the semi-classical Boltzmann transport equation (BTE) framework. Our study establishes the role of the strain-modulated electronic density of states, group velocities and energy-dependent scattering rates on the electronic transport properties. We highlight the role of different scattering processes, and how the predicted properties change when different electron scattering rate approximations, such as constant relaxation time and acoustic deformation potentials scattering approximations, are used. The physical insights demonstrated here not only are valuable to predict band movement and electron relaxation in complex heterostructures, but essential to develop strain engineering approaches to optimize electronic properties. 

\section{Methods}

\begin{figure}[h!]
\includegraphics[width=1.0\linewidth]{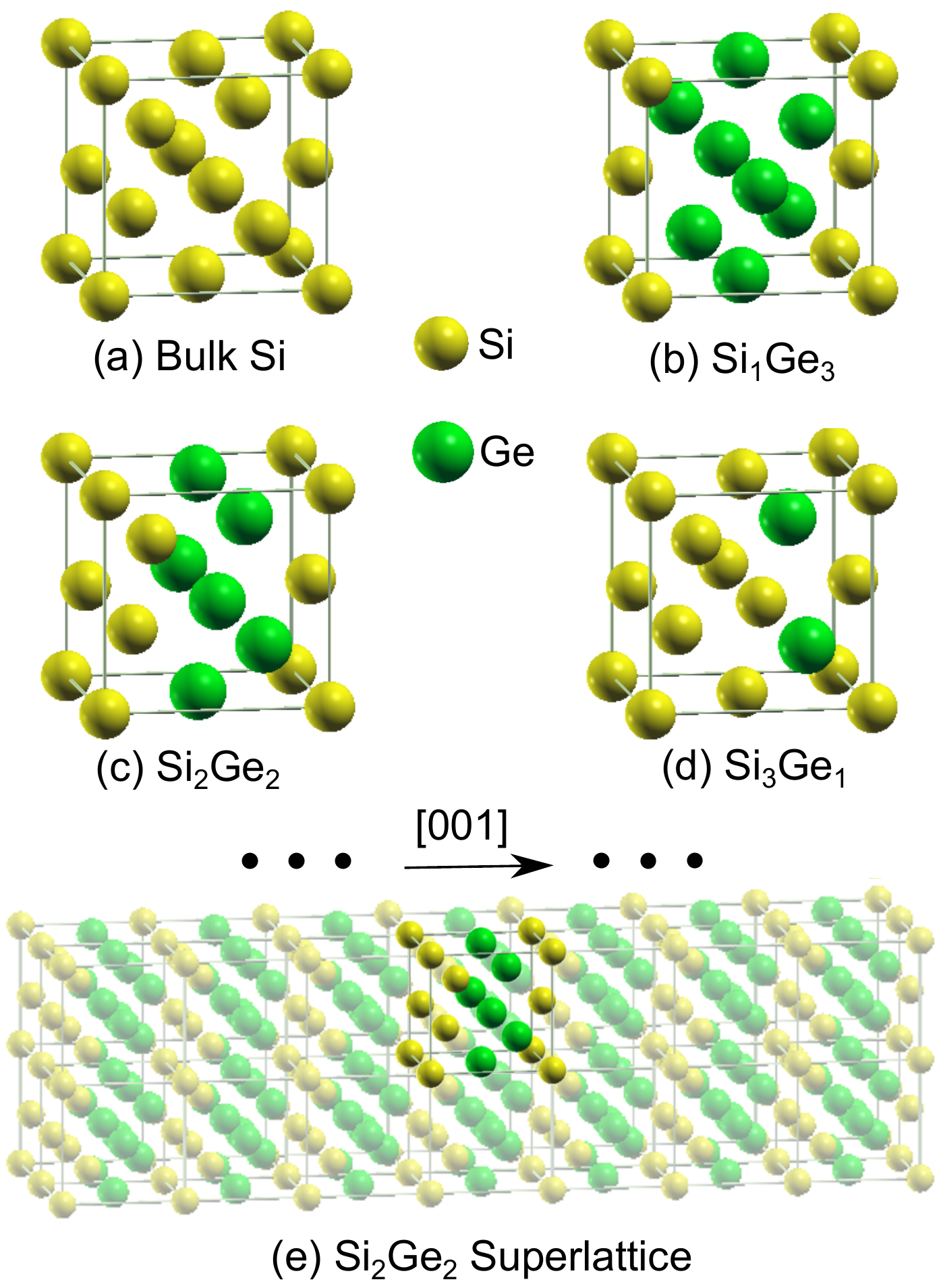}
\caption{\textbf{Representative configurations investigated:} (a) Eight-atom conventional unit cell of bulk Si. Eight-atom unit cells of superlattices: (b) Si$_1$Ge$_3$, (c) Si$_2$Ge$_2$, and (d) Si$_3$Ge$_1$. (e) Supercell of Si$_2$Ge$_2$ superlattice with periodic replicas along the [001] direction, shown with light shade.}
\label{fig:SLConfigs}
\end{figure}

Figure~\ref{fig:SLConfigs} shows the representative configurations of the model systems we investigate in this work: eight-atom Si$_\text{n}$Ge$_\text{m}$ SL unit cells composed of n and m monolayers of Si and Ge, respectively ($\text{n}+\text{m}=4$).  The configurations include atomically ordered electronic well (Si) and barrier (Ge) layers (Fig.~\ref{fig:SLConfigs}(e)), similar to other short-period SLs investigated in previous studies~\cite{zhang2013genetic,peter1996fundamentals}. We consider pristine SL models that are free of dislocations and other defects.  
We investigate the electronic transport properties of SL models with carrier concentration between $10^{17}$ and $10^{21}\ \text{cm}^{-3}$ at 300 K. In this doping and temperature range, EPS and IMS processes are the dominant mechanisms causing electron relaxation~\cite{li2012semiconductor,restrepo2009first,lundstrom2009fundamentals}. 

\subsection{Electronic Transport Coefficients}

We calculate the Seebeck coefficient (S), and the resistivity ($\rho$) of Si/Ge SLs following the BTE framework. We evaluate the integrals~\cite{ashcroftmermin,madsen2006boltztrap}
\begin{equation}
\mathcal{L}_{\parallel, \perp}^{(n)}(\text{E}_\text{F}, \text{T})=\int d\text{E}\:\Xi_{\parallel, \perp} (\text{E}) (\text{E}-\text{E}_\text{F})^{n}\left(-\frac{\partial f(\text{E,T})}{\partial \text{E}} \right),
\label{eq:Lintegral}
\end{equation} 
over the respective SL first Brillouin (BZ) zone. Here, $\parallel$ and $\perp$ refer to the in-plane and the cross-plane directions of the SLs, respectively. The $\mathcal{L}$-integrands depend on the transport distribution function (TDF), $\Xi (\text{E})$, the n$^{\text{th}}$ power of energy difference between energy, E, and the Fermi level, E$_\text{F}$, and the derivative of Fermi-Dirac distribution function $f$ at temperature T. The TDF is proportional to an energy dependent relaxation time, $\tau(\text{E})$, and an area integral, determined by the density of states (DOS) $\left ( \propto \oint_{\text{E}_{\bf k}=\text{E}} \frac{d\mathcal{A}}{|{\bf v_k}|} \right )$ weighted by the group velocity squared, ${\bf v}_{{\bf k},(\parallel, \perp)}^2$, and given by:
\begin{equation}
\Xi_{\parallel, \perp} (\text{E}) = \frac{\tau(\text{E})}{\hslash (2\pi)^3} \oint_{\text{E}_{\bf k}=\text{E}} \frac{d\mathcal{A}}{|{\bf v_k}|} ({\bf v}_{{\bf k},(\parallel, \perp)})^2.
\label{eq:TDF}
\end{equation}
We focus on the cross-plane ($\perp$) transport properties, and drop the subscripts ($\parallel$,$\perp$) in the subsequent discussions. The cross-plane transport coefficients are determined from the $\mathcal{L}$-integrals, and are given by:
\begin{align}
\sigma &= \mathcal{L}^{0}, \quad \rho=\sigma^{-1},\label{eq:sigma}\\ 
\text{S} &= \frac{1}{e\text{T}}\mathcal{L}^{1}/\mathcal{L}^{0}, \label{eq:seebeck}
\end{align}
where $e$ is electron charge. We compute $\tau(\text{E})$'s using first-principles approach and include them to obtain the transport coefficients, following the BTE framework as implemented in the BoltzTrap code~\cite{madsen2006boltztrap}. We calculate the room temperature normal (diffusive) transport coefficients~\cite{zhou2015ab} of eight-atom Si/Ge SLs at several carrier concentrations of interest.

\subsection{Energy Dependent Electron Relaxation Times}

We provide a brief summary of the current understanding of the EPS and the IMS processes, and discuss our approach to calculate the electron relaxation times in the SLs. We encourage the interested reader to consult the extensive literature for further study~\cite{ponce2016epw,noffsinger2010epw,li2012semiconductor}. We ignore the effects of electron-electron interactions, since past studies showed that these mechanisms do not alter the bulk Si mobility predictions by more than a few percent, at any doping level~\cite{restrepo2009first,meyer1987ionized}. 

\subsubsection{Electron-Phonon Scattering}

We compute the EPS rates employing different approximations, and discuss the effect of these approximations on the predicted rates and the electronic transport properties of the SLs. To obtain the {\em ab initio} EPS rates, we evaluate the first order electron-phonon matrix elements,
\begin{align}
g^{\nu}_{m n}(\textbf{k},\textbf{q})=\langle\psi_{m\textbf{k}+\textbf{q}}|\partial_{\textbf{q}\nu}U|\psi_{n\textbf{k}}\rangle.
\label{eq:elphmatrix}
\end{align}
Here, $\psi_{n\textbf{k}}$ is the electronic wave function of the Kohn-Sham state with eigenenergy $\text{E}_{n\textbf{k}}$ corresponding to band $n$ and wavevector $\textbf{k}$, $\omega_{\textbf{q}\nu}$ is the frequency of the phonon mode with wavevector $\textbf{q}$ and branch $\nu$, and $U$ is the Kohn-Sham self-consistent electron-ion potential. We compute the electron and phonon eigenstates using DFT and the electron-phonon matrix elements, $g^{\nu}_{m n}(\textbf{k},\textbf{q})$, using DFPT~\cite{noffsinger2010epw,ponce2016epw}, on a coarse grid that samples the respective reciprocal spaces.  We then obtain interpolated electron and phonon eigenstates and $g^{\nu}_{m n}(\textbf{k},\textbf{q})$, on a dense grid, using maximally localized Wannier functions as implemented in the EPW package, which provides accurate interpolation of electron-phonon matrix elements from coarse grids to dense grids~\cite{park2014electron}. The Wannier90 package within EPW, which performs these interpolations, was demonstrated to successfully predict the electron-phonon coupling and superconducting properties of several solid state systems~\cite{noffsinger2010epw,ponce2016epw}. This approach, in conjunction with exact solution of linearized BTE, has shown to accurately predict bulk Si's resistivities and Seebeck coefficients matching measured data~\cite{fiorentini2016thermoelectric}. Using the dense grid electron-phonon matrix elements, $g^{\nu}_{m n}(\textbf{k},\textbf{q})$, we obtain the electron self energy ($\Sigma$): \begin{align}
\Sigma_{n\textbf{k}}&=\frac{1}{\hslash}\sum_{m\nu}\int\frac{d\textbf{q}}{\Omega_{BZ}}|g^{\nu}_{m n}(\textbf{k},\textbf{q})|^2\nonumber\\ &\times\bigg\{\left[\frac{n(\omega_{\textbf{q}\nu},\text{T})+f(\text{E}_{m\textbf{k}+\textbf{q}}/\hslash, \text{T})}{\text{E}_{n\textbf{k}}/\hslash-\text{E}_{m\textbf{k}+\textbf{q}}/\hslash-\omega_{\textbf{q}\nu}+i\eta}\right]\nonumber\\ &+\left[\frac{n(\omega_{\textbf{q}\nu},\text{T})+1-f(\text{E}_{m\textbf{k}+\textbf{q}}/\hslash, \text{T})}{\text{E}_{n\textbf{k}}/\hslash-\text{E}_{m\textbf{k}+\textbf{q}}/\hslash+\omega_{\textbf{q}\nu}+i\eta}\right]\bigg\}.
\label{eq:elecselfE}
\end{align}
Here $\Omega_{BZ}$ is the BZ volume, $n(\omega_{\textbf{q}\nu},\text{T})$ and $f(\text{E}_{m\textbf{k}+\textbf{q}}/\hslash, \text{T})$  are the Bose-Einstein and Fermi-Dirac occupation functions for phonons and electrons at temperature T, respectively. 
The real and imaginary parts of self energy~\cite{ponce2016epw,giustino2017electron} are given by $\Sigma_{n\textbf{k}}'$ and $\Sigma_{n\textbf{k}}''$, respectively: $\Sigma_{n\textbf{k}}=\Sigma_{n\textbf{k}}'+i\Sigma_{n\textbf{k}}''$.
We obtain the electron-phonon relaxation times ($\tau_\text{ep-EPW}$) from the imaginary parts of the self energy: 
\begin{align}
\frac{1}{\tau_{\text{ep-EPW},n\textbf{k}}}=\frac{2|\Sigma_{n\textbf{k}}''|}{\hslash}.
\label{eq:elphrelax}
\end{align}

The occupation functions in Eq.~\ref{eq:elecselfE} introduce the effect of temperature on electron relaxation times. However, we assume that the lattice vibrations are in thermal equilibrium during the scattering process. This assumption is not strictly valid when a temperature gradient is present across a material. The temperature gradient could establish an additional electrical current in the same direction as the heat flow, resulting in a phonon drag contribution to the Seebeck effect that boosts the diffusive part. The effect of such coupled electron-phonon dynamics on the thermoelectric properties of silicon recently received renewed attention~\cite{zhou2015ab,mahan2014seebeck,fiorentini2016thermoelectric}. Several phonon engineering strategies have been proposed to improve S with nanostructuring~\cite{fiorentini2016thermoelectric,nadtochiy2019enhancing}, however, contradictory results have been reported~\cite{boukai2008silicon,sadhu2015quenched}. To accurately account for the effect of phonon engineering approaches on electronic transport properties, one needs to solve the coupled BTE's for electrons and phonons, which makes it especially challenging for SL nanostructures, where phonons are strongly affected and might deviate from bulk character. We believe that our predicted trends of thermoelectric properties of SLs with diverse lattice strain environment will persist after considering this effect. We leave the discussion of phonon drag contributions to the SL thermoelectric properties for future study. We also did not include the direct effect of doping on EPS rates, and assumed that electrons are in their equilibrium energy states. Previous {\em ab initio} study of SiGe compound showed that electron relaxation times are relatively unaffected by change in carrier concentration, as opposed to phonon relaxation times which can be significantly reduced by electron–phonon coupling at high carrier concentrations~\cite{fan2018first}. 

We compare the {\em ab initio} EPW predicted EPS rates with the rates obtained considering acoustic deformation potential scattering in the elastic limit. In this limit, the EPS rates in non-polar semiconductors are assumed to generally follow the DOS~\cite{fischetti1991monte}. We refer to this concept as DOS scattering. We exploit the proportionality to predict the relaxation times within the DOS scattering approximation as $\tau_\text{ep-DOS}$: $1/\tau_\text{ep-DOS}(\text{E}) = \text{K}_\text{el-ph}\times\text{DOS}(\text{E})$. We apply the DOS scattering approximation near the conduction band edge, the region of interest for transport properties of n-doped SLs. We obtain the scaling constants, K$_{\text{el-ph}}$, by aligning DOS(E) with the EPS-EPW rates. The proportionality constant, K$_{\text{el-ph}}$, is related to the deformation potential squared. Recently, deformation potential based concepts have been utilized to develop computationally efficient approaches (AMSET) to predict EPS rates~\cite{ganose2021efficient}. In bulk systems like Si (Ge), electron transport properties are mainly determined by the $\Delta$ (L) valley electrons, and the deformation potentials can successfully describe the properties of the states within these valleys. Subsequently, the DOS scattering approximation has been exploited to accurately predict the electron and hole mobilities of bulk silicon~\cite{witkoske2017thermoelectric,yu2008first}. In an earlier study, we computed thermopower (S) of Si/Ge SLs employing DOS scattering approximation (see Fig. S2, in Ref.~\citenum{proshchenko2019optimization}), and compared with S obtained using CRTA. We noted that the trends of S obtained with the two approximations match quite well, although the values differ from each other. However, we acknowledged that a detailed analysis of the validity of the DOS scattering approximation would be highly beneficial since this approach could significantly expedite the computation of EPS rates. In this article, we discuss the applicability of DOS approximation to predict energy-dependent EPS rates and resulting electronic transport properties of Si/Ge SLs. We demonstrate that it is not straightforward to apply such approximation, especially since the value of K$_{\text{el-ph}}$ is highly dependent on the fitting procedure. We include figures demonstrating the variability of K$_{\text{el-ph}}$ values and the effect on the predicted electronic transport properties in the Supplementary Information (SI). We conclude that further work is required to substitute the expensive EPS-EPW approach with the DOS scattering approximation approach. 

\subsubsection{Electron-Ionized Impurity Scattering}

In addition to electron-phonon coupling, electron relaxation in Si/Ge SLs is strongly affected by scattering due to ionized impurities, which are ubiquitous in doped semiconductors~\cite{chattopadhyay1981electron}. We employ a modified Brooks-Herring (BH) formulation to model the electron-ionized impurity scattering mechanisms. In the BH model, the ionized impurities are assumed to be static, randomly spaced objects, and a spherically symmetric Yukawa potential is used to model the screening effect on electrons. This model provides an improved representation of real systems by explicitly including a long-range Coulomb tail for a screened ionized impurity, which was missing in the bare Coulomb potential form of the Conwell and Weisskopf's model~\cite{conwell1950theory}. The impurities are assumed to not have any internal excitations, so the electron scatters from them elastically. The impurity concentration is assumed to be dilute, therefore, interference between successive scatterings is neglected. We implement this simplifying assumption to reduce the complexity of the problem, however, we acknowledge that this assumption may not be strictly valid at higher carrier concentrations. The electron states are Bloch waves except for occasional scattering from isolated impurities. The impurity causes the electron in state $\textbf{k}$ and band $n$ to elastically scatter to state $(\textbf{k}+\textbf{q}-\textbf{G})$ and band $m$, which has the same energy. Here, $\textbf{G}$ is a reciprocal lattice vector. We use Fermi's golden rule to calculate the electron relaxation time due to impurity scattering as~\cite{li2012semiconductor,restrepo2009first}:
\begin{align}
\frac{1}{\tau_{im,n\textbf{k}}}&=\frac{2\pi \text{n}_\text{i}}{\hslash}\frac{V^2}{(2\pi)^3}\sum_\textbf{G}\sum_m  \int d\textbf{q}\: H^2_{n\textbf{k},m\textbf{k}+\textbf{q}-\textbf{G}}\nonumber\\
&\times\delta(\text{E}_{n\textbf{k}}-\text{E}_{m\textbf{k}+\textbf{q}-\textbf{G}})(1-\cos{\theta}).
\label{eq:imprelax}
\end{align}
Here, $\text{n}_\text{i}$ is the density of scattering centers. We assume that all dopants in our models are ionized at $\text{T}=300$K. Therefore, the density of scattering centers ($\text{n}_\text{i}$) is same as the carrier concentration ($\text{n}_\text{e}$). $V$ is the volume of the SL configurations, $\theta$ is the angle between electron group velocities in state $\textbf{k}$ and band $n$, $\textbf{v}_{n\textbf{k}}$, and in state $\textbf{k}'=\textbf{k}+\textbf{q}-\textbf{G}$ and band $m$, $\textbf{v}_{\textbf{k}+\textbf{q}-\textbf{G}}$, before and after scattering, respectively. The $\textbf{G}=0$ terms correspond to \textit{normal} scattering and the $\textbf{G}\neq 0$ to \textit{umklapp} scattering~\cite{jacoboni1983monte}, respectively. 

The matrix element, $H_{n\textbf{k},m\textbf{k}+\textbf{q}-\textbf{G}}$, represents the perturbation term of the electron-ionized impurity interaction Hamiltonian. We model the perturbation using an anisotropic impurity scattering potential, $\phi(\textbf{r})$. We derive the anisotropic potential, $\phi(\textbf{r})$, by solving the screened Poisson-Boltzmann equation, given by
\begin{align}
\nabla\cdot(\mbox{\boldmath$\varepsilon$}\nabla\ - \lambda_D^{-2})\phi(\textbf{r}) = \frac{-e\delta^3(\textbf{r})}{\varepsilon_0}, \hspace{0.2in} \mbox{\boldmath$\varepsilon$} = \begin{bmatrix}
\varepsilon_\parallel & 0 & 0\\
0 & \varepsilon_\parallel & 0\\
0 & 0 & \varepsilon_\perp
\end{bmatrix}.
\label{eq:PoissonBoltzmann}
\end{align}
Here, $\lambda_D$ is the modified Debye screen length, $\lambda_D=\sqrt{\varepsilon_0 k_{B}\text{T}/e^2\text{n}_\text{i}}$, $k_{B}$ is the Boltzmann constant and $\varepsilon_0$ is the dielectric permittivity of vacuum. We consider an anisotropic dielectric tensor, $\mbox{\boldmath$\varepsilon$}$, consisting of different in-plane ($x-y$) ($\varepsilon_\parallel$) and cross-plane ($z$) ($\varepsilon_\perp$) dielectric constants, to explicitly account for the in-plane and cross-plane structural anisotropy of the SL configurations, respectively. The in-plane SL structural symmetry further imposes, $\varepsilon_{xx}=\varepsilon_{yy}=\varepsilon_\parallel$. We solve for $\phi(\textbf{r})$ from Eq.~\ref{eq:PoissonBoltzmann}, by taking Fourier transform and performing a transformation of variables from ($\textbf{r}$ and $\textbf{k}$) to ($\textbf{r}^\prime$ and $\textbf{k}^\prime$), where $(r_x',r_y',r_z')=(r_x/\sqrt{\varepsilon_\parallel},r_y/\sqrt{\varepsilon_\parallel},r_z/\sqrt{\varepsilon_\perp})$ and $(k_x',k_y',k_z')=(k_x\sqrt{\varepsilon_\parallel},k_y\sqrt{\varepsilon_\parallel},k_z\sqrt{\varepsilon_\perp})$. The solution, in the form of a Yukawa potential, is given by 
\begin{align}
\phi(\textbf{r}) = \frac{e}{4\pi r' \varepsilon_0 \sqrt{\varepsilon_\parallel^2\varepsilon_\perp}}e^{-r'/\lambda_D},
\label{eq:imppotential}
\end{align}
where $r'=|\textbf{r}^{\prime}|=\sqrt{(r_x^2+r_y^2)/\varepsilon_\parallel+r_z^2/\varepsilon_\perp}$. We obtain the dielectric constants $\varepsilon_\parallel$ and $\varepsilon_\perp$ using first-principles calculations, as described in the next subsection, and compute $\phi(\textbf{r})$ numerically for the subsequent analysis. We use the anisotropic potential, $\phi(\textbf{r})$, to obtain the perturbation matrix elements, $H_{n\textbf{k},m\textbf{k}+\textbf{q}-\textbf{G}}$, from
\begin{align}
H_{n\textbf{k},m\textbf{k}+\textbf{q-G}}&=\frac{1}{V}\int d\textbf{r}\psi^*_{m\textbf{k}+\textbf{q}-\textbf{G}}(\textbf{r})\phi(\textbf{r})\psi_{n\textbf{k}}(\textbf{r}),
\label{eq:impmatrixelInt}
\end{align}
where the electronic wave functions $\psi_{n\textbf{k}}$ are given by, $\psi_{n\textbf{k}} = 1/\sqrt{V}u_{n\textbf{k}}(\textbf{r})e^{i\textbf{k}\cdot\textbf{r}}$. $u_{n\textbf{k}}(\textbf{r})$'s are the periodic parts of the Bloch states. Inserting the expression of $\phi(\textbf{r})$ from Eq.~\ref{eq:imppotential} into Eq.~\ref{eq:impmatrixelInt} and performing the Fourier integral over the crystal volume, we obtain  
\begin{align}
H_{n\textbf{k},m\textbf{k}+\textbf{q-G}}&=\frac{e^2 \lambda_D^2}{V\varepsilon_0 (1+|\textbf{q}'|^2\lambda_D^2)}\zeta_{n\textbf{k},\textbf{k}+\textbf{q}-\textbf{G}},\label{eq:impmatrixel} \\
\text{with} \ \zeta_{n\textbf{k},m\textbf{k}+\textbf{q}-\textbf{G}}&=\int d\textbf{r}u^*_{m\textbf{k}+\textbf{q}-\textbf{G}}(\textbf{r})u_{n\textbf{k}}(\textbf{r})e^{i\textbf{G}\cdot\textbf{r}},
\label{eq:overlap}
\end{align}
where $\textbf{q}'=(\sqrt{\varepsilon_\parallel}\textbf{i}+\sqrt{\varepsilon_\parallel}\textbf{j}+\sqrt{\varepsilon_\perp}\textbf{k})\cdot\textbf{q}$. Here, $\zeta_{n\textbf{k},m\textbf{k}+\textbf{q}-\textbf{G}}$ is the overlap integral, which we assume to be unity~\cite{fiorentini2016thermoelectric,deng2020epic}. We compute the IMS rates using the matrix elements, $H_{n\textbf{k},m\textbf{k}+\textbf{q}-\textbf{G}}$, following Eq.~\ref{eq:imprelax}.

Below, we include a brief discussion about two of the approximations we made to calculate the IMS rates. First is the approximation that considers the overlap integral to be unity. We performed rigorous evaluations of the overlap integrals using the Wannier functions and calculated the resultant scattering rates and the electronic transport coefficients. We noted that the full considerations of $\zeta_{n\textbf{k},m\textbf{k}+\textbf{q}-\textbf{G}}$ do not significantly affect the electronic transport properties at most carrier concentrations of interest. However, the computational cost increases by $\sim~3-4$-fold compared to the cases that assumed $\zeta_{n\textbf{k},m\textbf{k}+\textbf{q}-\textbf{G}}=1$, for different carrier concentrations. The predicted S is considerably less affected due to the inclusion of the overlap factors compared to $\rho$. For example, the two S predictions stay within 5\% for Si$_1$Ge$_3$ for all $n_e$ of interest and maximum deviation of S is about 15\% above $n_e=10^{20}$ cm$^{-3}$ (Si), and $7.5\times 10^{20}$ cm$^{-3}$ (Si$_2$Ge$_2$). However, such incorporation results in nonuniform variations of predicted resistivities in the high carrier concentration regime. We did not observe a definite trend of the deviations of SL transport coefficients by including the overlap factors. We acknowledge that this aspect needs a thorough investigation, however, leave it for future work and focus on the discussion of the trends of transport properties of SL systems. Additionally, we only include contributions from the $G = 0$ normal scattering terms and ignore the $G \neq 0$ umklapp terms, due to the fact that $H_{n\textbf{k},m\textbf{k}+\textbf{q}-\textbf{G}}$ decreases with increasing $|\textbf{q}'|$, as can be seen from Eq.~\ref{eq:impmatrixel}. A number of umklapp terms could contribute in principle, however, normal scattering terms are expected to play a dominant role to determine the EMS rates~\cite{jacoboni1983monte}. We acknowledge that umklapp terms may be comparable to the normal terms for SL systems, however, further studies are required to establish the relative importance of the two terms. We do not include these terms to keep the computational expenses at minimum. 

We average the $\textbf{k}$-dependent scattering rates for each type of scattering, $I=\text{ep},\:\text{im}$, to obtain the energy-dependent electron relaxation times. We numerically evaluate $\tau_{I}(\text{E})$ from
\begin{align}
\frac{1}{\tau_\text{ep-EPW}(\text{E})}&= \frac{1}{\text{DOS}(\text{E})}\sum_n  \int_{\text{BZ}} \frac{d\textbf{k}}{\Omega_{\text{FBZ}}}\frac{1}{\tau_{ep-EPW,n\textbf{k}}}\delta(\text{E}-\text{E}_{n\textbf{k}}), \nonumber \\
\frac{1}{\tau_\text{im}(\text{E})}&= \frac{1}{\text{DOS}(\text{E})}\sum_n  \int_{\text{BZ}} \frac{d\textbf{k}}{\Omega_{\text{FBZ}}}\frac{1}{\tau_{im,n\textbf{k}}}\delta(\text{E}-\text{E}_{n\textbf{k}}),
\label{eq:tau-energy}    
\end{align}
by averaging $1/\tau_{I,n\textbf{k}}$, corresponding to $\text{E}_{n\textbf{k}}$ in the energy window $\text{E}\pm \Delta$ with $\Delta=5\times10^{-4}$ Ry. Here $\Omega_{\text{FBZ}}$ is the FBZ volume. We provide the relevant numerical details in the following Computational Details section. This numerical averaging procedure has been demonstrated to successfully predict the energy-dependent relaxation times in several studies~\cite{deng2020epic}, including superconducting cuprates that have highly anisotropic cell-axes and conduction properties~\cite{deng2020epic,schulz1992hall,samsonidze2018accelerated,allen1988anisotropic,madsen2006boltztrap}. We calculate the total electron relaxation times due to electron-phonon ($\text{ep}$) and electron-ionized impurity ($\text{im}$) scattering processes, using Matthiessen's rule~\cite{deng2020epic}:
\begin{align}
\frac{1}{\tau(\text{E})}&=\frac{1}{\tau_\text{ep-EPW}(\text{E})} + \frac{1}{\tau_\text{im}(\text{E})}, \ \ \text{  and} \\
\frac{1}{\tau(\text{E})}&=\frac{1}{\tau_\text{ep-DOS}(\text{E})} + \frac{1}{\tau_\text{im}(\text{E})}.
\label{eq:matthiessen}
\end{align}

\subsection{Computational Details}

\subsubsection{Energy Dependent Electron Relaxation Times}

We calculate the electronic structure properties of eight-atom Si/Ge SLs by performing self-consistent field (SCF) and non self-consistent field (NSCF) DFT calculations, as implemented in the plane-waves code Quantum Espresso (QE)~\cite{giannozzi2009quantum}. We use non-relativistic norm-conserving pseudopotentials for Si and Ge where the valence electrons ($3s^23p^2$) are treated with the Perdrew-Zunger parametrization of the local-density approximation (LDA) of the exchange–correlation energy functional~\cite{perdew1981self}. Similar numerical approach has been followed in previous DFT studies that computed electron scattering rates in Si~\cite{ponce2016epw,fiorentini2016thermoelectric,fu2017electron}. We use cutoff energy of 45 Ry to expand the Kohn-Sham orbitals in terms of a plane wave basis set for all calculations. A convergence threshold of $10^{-9}$ Ry is chosen for self-consistency. We ignore the effects of spin-orbit coupling on the energy bands in our calculations. The LDA approximation is known to under-predict bulk Si and Ge bandgaps. We compute the band gaps of the Si/Ge SLs separately using the Heyd-Scuseria-Ernzerhof (HSE06) hybrid functionals~\cite{heyd2003hybrid}. We use a scissors operator to correct the LDA band gaps using the HSE06 predictions, shown in Table~\ref{tab:dielectric}. We provide the details regarding geometry optimization and HSE06 band gap calculations in SI. 

\begin{figure} [htbp]
\includegraphics[width=1.0\linewidth]{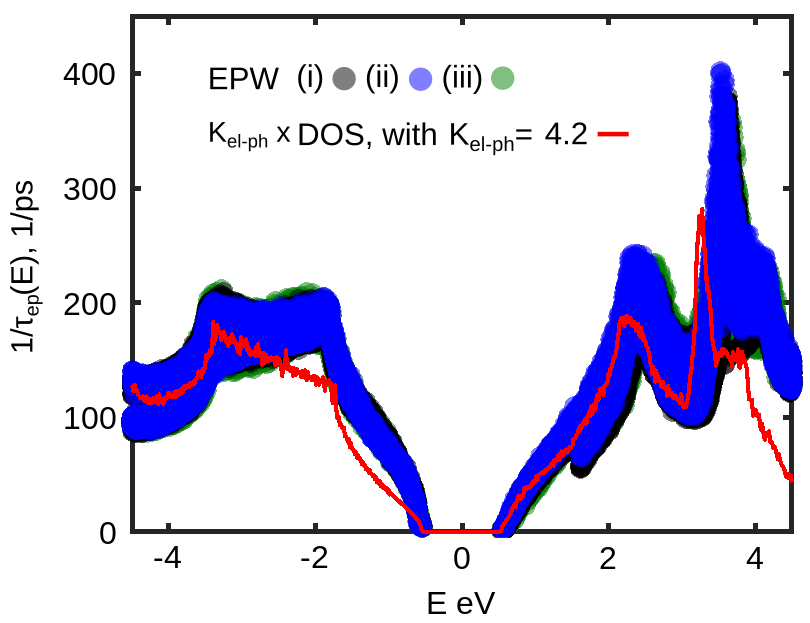}
\caption{Electron-phonon scattering rates in (i) two-atom primitive cell (grey) and (ii) eight-atom conventional cell (blue) of bulk Si. (iii) Reference data (green) is extracted from Ref.~\citenum{ponce2016epw}. DOS scattering rates in eight-atom conventional cell are shown with red line. We obtain identical EPS-DOS rates for two-atom primitive cell, confirming that DOS scattering rates are independent of the unit cell sizes for bulk Si.}
\label{fig:bulkSi_scattering_rate}
\end{figure}

We perform NSCF calculations to obtain electronic states and DFPT calculations to obtain phonon modes. We use a coarse grid sampling of the respective Brillouin zones to compute electronic and phonon states and electron-phonon matrix elements, $g^{\nu}_{m n}(\textbf{k},\textbf{q})$. We use Wannier functions to interpolate $g^{\nu}_{m n}(\textbf{k},\textbf{q})$ on a fine grid using the EPW code~\citet{ponce2016epw}. We compute the EPS rates at T$=$ 300 K using the interpolated $g^{\nu}_{m n}(\textbf{k},\textbf{q})$, by following the steps outlined in Eqs.~\ref{eq:elphmatrix}$-$\ref{eq:elphrelax} of the previous subsection. We first calculate the EPS rates in two-atom primitive unit cell of undoped bulk Si and compare with literature data~\cite{ponce2016epw}. This step establishes confidence in our numerical approach to produce results consistent with published data. We perform SCF and NSCF calculations to compute the electronic states and DFPT calculations to compute the phonon modes of bulk Si. We calculate the electronic states, phonon modes, and electron-phonon matrix elements on a $6\times6\times6$ coarse Monkhorst-pack grid~\cite{monkhorst1976special}. We then obtain interpolated electron-phonon matrix elements on a fine grid of $30,000/150,000$ randomly selected k/q-points, using the EPW code~\cite{ponce2016epw}. We choose these coarse and fine grid sizes to compare our results with literature data, obtained with similar coarse (fine) $6\times6\times6$/$6\times6\times6$ (30,000/150,000) k/q-points grids~\cite{ponce2016epw}. We average $\tau_{n\textbf{k}}$ over a $\sim 30,000$ k-mesh to obtain energy dependent electron relaxation times due to phonons, $\tau_\text{ep-EPW}(\text{E})$. We show the resulting EPW-EPS rates in two-atom bulk Si models in Fig.~\ref{fig:bulkSi_scattering_rate} (grey circles). We use a Gaussian broadening of $10$ meV, and the HSE06 predicted bulk Si band gap of 1.08 eV to adjust the EPS-EPW rates. The green circles in Fig.~\ref{fig:bulkSi_scattering_rate} represent the reported EPW predicted scattering rates in two-atom bulk Si models, extracted from Ref.~\citenum{ponce2016epw}. Note that the literature data are adjusted to display the Si band gap, 1.08 eV, used for this article. The red line corresponds to EPS rates obtained with the DOS scattering approximation, $\tau_\text{ep-DOS}$: $1/\tau_\text{ep-DOS}(\text{E}) = \text{K}_\text{el-ph}\times\text{DOS}(\text{E})$, with $\text{K}_\text{el-ph}=4.2$. We use a dense $\sim 200,000$ k-mesh for NSCF calculations and an energy step of $0.001$ Ry to obtain DOS(E). The DOS has the units of number of states/Ryd/spin/unit cell, henceforth referred to as DOS.u.. The comparison between the EPS-EPW and EPS-DOS rates shows that the DOS approximation is quite effective to describe EPS near the conduction band edge. However, the EPS-DOS predictions deviate as we move away from the edge.

To find the appropriate coarse and fine grid sizes for the SLs, we perform extensive tests to compute EPS rates in a similar size eight-atom bulk Si conventional unit cell, and compare these results with the {\em reference} two-atom Si EPS rates~\cite{ponce2016epw}. The use of grid sizes, similar to the above mentioned 2-atom calculations, to obtain the EPS rates in several eight-atom Si/Ge SLs, demands unfeasible computational expenses. Additionally, such sampling might not be necessary for the respective eight-atom BZs, which is four times smaller than the primitive cell BZ. Figure~\ref{fig:bulkSi_scattering_rate} shows the comparison between EPS rates in eight-atom bulk Si conventional unit cell (blue), two-atom primitive cell, and the reference data (green). The eight-atom EPS rates, obtained with coarse (fine) $6\times6\times6$/$6\times6\times6$ (15,000/50,000) k/q-points grids match well with the two-atom results and the reference data, obtained with coarse (fine) $6\times6\times6$/$6\times6\times6$ (30,000/150,000) k/q-points grids. We average $\tau_{n\textbf{k}}$ over a $\sim 15,000$ k-mesh for eight-atom bulk Si, to obtain energy dependent EPS rates. We then systematically vary the coarse and the fine BZ sampling grids to obtain converged results for the eight-atom EPS rates. This aspect is particularly important since BZ sampling significantly impacts the computed transport properties. The importance of grid size based convergence tests has been acknowledged in multiple studies~\cite{ponce2018towards}. We find that EPS rates obtained with $6\times6\times6$/$3\times3\times3$ (15,000/50,000) coarse (fine) k/q-point grids match well with $6\times6\times6$/$6\times6\times6$ (15,000/50,000) coarse (fine) k/q-point grid results.

Such dense sampling proves to be expensive for calculating EPS rates in SLs with a diverse lattice strain environments. We check further to find an appropriate BZ sampling of SLs that produces converged EPS rates, and also helps to keep the computational costs manageable. We compute EPS rates in Si$_2$Ge$_2$ SL, assumed to be grown on Si, with (coarse/fine) $4\times4\times4$/$4\times4\times4$ (10,000/30,000) k/q-point grids. The rates match well with the coarse (fine) $6\times6\times6$/$3\times3\times3$ (15,000/50,000) k/q-points grid results for Si$_2$Ge$_2$ SL on Si. {\em We use the coarse (fine) $4\times4\times4$/$4\times4\times4$ (10,000/30,000) BZ sampling to calculate all the Si/Ge SL EPS rates presented in this article.} We average $\tau_{n\textbf{k}}$ over a $\sim 10,000$ k-mesh to obtain energy dependent electron-phonon relaxation times in eight-atom SLs. We report the results from our extensive convergence tests in SI for the benefit of the readers interested in reproducing our results. In order to compute IMS relaxation times, $\tau_\text{im}(\text{E})$, we perform NSCF calculations using a dense $\sim 200,000$ k-mesh, and obtain dielectric constant tensors using DFPT calculations as implemented in QE~\cite{giannozzi2009quantum}. We implement the random phase approximation including local field effects (LRPA) to obtain the dielectric constants. We list the LRPA calculated anisotropic dielectric constants and the HSE06 predicted band gaps of the Si/Ge SLs used in our calculations in Table~\ref{tab:dielectric}. We find that the LRPA predicted dielectric constant of bulk Si (12.66) is greater than that of the measured value of 11.9~\cite{dunlap1953direct} by $\sim6.4\%$, however, matches well with previous GW+BSE result (12.7)~\cite{zheng2017frequency}. We combine $\tau_\text{ep-EPW}(\text{E})$ and $\tau_\text{im}(\text{E})$ to calculate $\tau(E)$ using Matthiessen's rule,  (Eq.~\ref{eq:matthiessen}) as described in the previous subsection.

\begin{table}
\caption{\label{tab:dielectric}In-plane and cross-plane diagonal components of the anisotropic dielectric tensors and HSE06 computed band gaps of the Si/Ge SLs.}
\begin{tabular}{ccccc}
\hline
System&Substrate&$\varepsilon_\parallel$&$\varepsilon_\perp$&Band gap (HSE06)\\
\hline
Si$_1$Ge$_3$&Si&14.38&14.87&0.49 eV\\
Si$_2$Ge$_2$&Si&13.66&13.95&0.55 eV\\
Si$_3$Ge$_1$&Si&12.87&13.05&0.69 eV\\
Si$_2$Ge$_2$&Si$_{0.5}$Ge$_{0.5}$&14.11&13.83&0.80 eV\\
Si$_2$Ge$_2$&Ge&14.95&13.99&0.63 eV\\
\end{tabular}
\end{table}

\begin{figure}[htbp]
\includegraphics[width=1.0\linewidth]{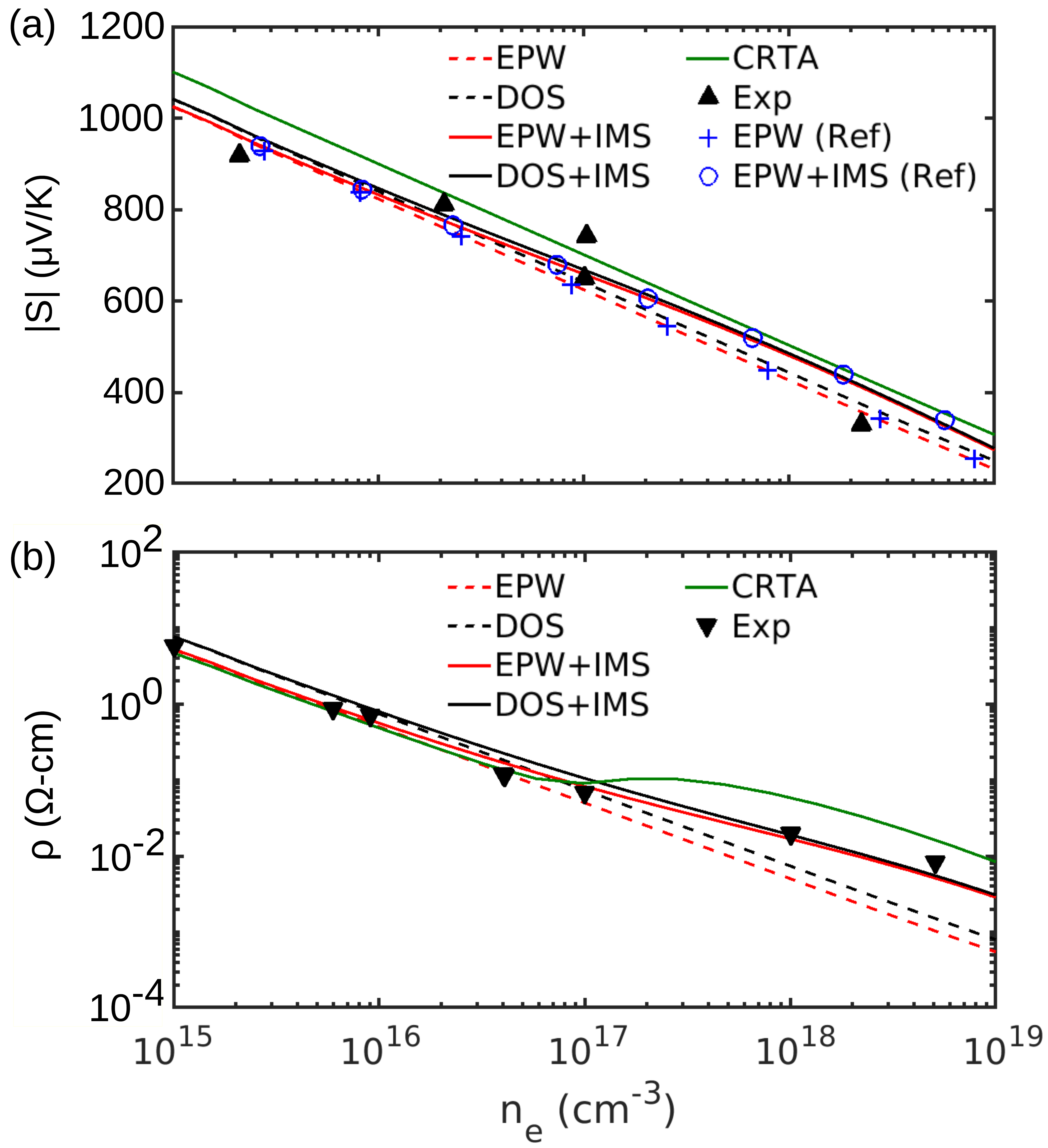}
\caption{Comparison of predicted (a) Seebeck coefficient (S) and (b) resistivity ($\rho$) of bulk Si and literature data. The dashed red and black lines represent EPS-EPW and EPS-DOS results, respectively. The solid red and black lines represent predictions made considering EPS-EPW+IMS and EPS-DOS+IMS, respectively. The solid green line represent results obtained using the CRTA approximation. The difference between the lines show how S and $\rho$ predictions vary for different scattering approximations. For CRTA calculations, $\tau$ is extracted from the measured bulk Si mobility~\cite{jacoboni1977review}. The black triangles represent measured data for diffusive S, extracted from Refs.~\citenum{geballe1955seebeck, herring1954theory}. The blue +'s (EPS-EPW) and blue circles (EPS-EPW+IMS) represent past first-principles study predictions of diffusive S, for comparison. These values are extracted from Ref.~\citenum{fiorentini2016thermoelectric}. The measured data for $\rho$ are extracted from Ref.~\citenum{mousty1974relationship} (inverted black triangles).}
\label{fig:bulkSi_resistivity}
\end{figure}

\subsection{Electronic Transport Coefficients}

We calculate the electronic transport coefficients at several carrier concentrations of interest, by incorporating the energy dependent relaxation times, $\tau(\text{E})$. We consider different scattering approximations to obtain $\tau$, and compare the resulting transport properties. We obtain TDFs using the BoltzTrap code which considers $\tau(\text{E})=1$ for all energies. With our in-house code, we incorporate $\tau(\text{E})$ with the TDFs, and compute S and $\rho$ following steps outlined in Eqs.~\ref{eq:Lintegral}-\ref{eq:seebeck}. We use an energy step of 0.001 Ry to compute the BTE integrals. We use the HSE06 corrected band gap for bulk Si (1.08 eV) for both S and $\rho$ calculations, as opposed to the LDA predicted band gap used in past research~\cite{fiorentini2016thermoelectric}. Figure~\ref{fig:bulkSi_resistivity} shows the comparison of the predicted (a) Seebeck coefficient (semi-log plot) and (b) the resistivity (log-log plot) of eight-atom bulk Si with literature results. The Seebeck coefficients predicted with $\tau_\text{ep-EPW}$ and $\tau_\text{ep-DOS}$ are shown with red and black dashed lines in Fig.~\ref{fig:bulkSi_resistivity}(a), respectively. The match between the two results suggests that the DOS scattering approximation works well to describe the bulk Si Seebeck coefficients. The blue pluses are previous first-principles EPS-EPW results extracted from Ref.~\citenum{fiorentini2016thermoelectric}. The data represent diffusive S predicted using a \textbf{k}-dependent scattering approach and exact BTE formulation. We find that our S predictions made considering an energy-dependent electron relaxation time, $\tau(\text{E})$, compare well with exact BTE based predictions. The solid red and black lines represent predictions made considering total electron relaxation times due to EPS-EPW+IMS and EPS-DOS+IMS, respectively. The corresponding literature results, extracted from Ref.69, are shown with blue circles. The close match between the two sets of results can be noted. Past studies reported that the diffusive part of S does not depend strongly on the nature of the scattering mechanisms and switching off phonon or impurity scattering affects the S result by less than 8\%~\cite{fiorentini2016thermoelectric}. We also note that the predicted S results, including the IMS rates (solid lines), do not show considerable deviations from the results with only EPS rates included (dashed lines). The maximum deviation of 18\% between the two results can be observed at n$_\text{e}\sim10^{19}$ cm$^{-3}$. The black triangles represent measured data for diffusive contributions to S, and are extracted from Refs.~\citenum{geballe1955seebeck, herring1954theory}. We find that our results compare well with the measured data, especially in the low doping regime. At high doping, for n$_\text{e}$\textgreater$10^{17}$cm$^{-3}$, our results are slightly higher than the experimental data. The discrepancy between theoretical predictions and measured data was attributed to possible electron-plasmon interactions~\cite{fiorentini2016thermoelectric}. The solid green line represent results obtained using $\tau$ within the CRTA approximation. We calculate the TDFs and consider doping-dependent relaxation times, $\tau(\text{n}{_\text{e}})$, to calculate S. We use the $\tau(\text{n}{_\text{e}})$ of unstrained bulk Si, parametrized by Ref.~\cite{hinsche2012thermoelectric} using the values first reported in Ref.~\cite{jacoboni1977review}. We find that CRTA predicted S is higher than the EPS-EPW+IMS results by atmost 10\%, indicating the relative independence of S on scattering mechanisms. 
Similar observation can be made by noting the match between Seebeck coefficients predicted using EPW approach and the DOS approximation. 

We show the bulk Si resistivity predicted using $\tau_\text{ep-EPW}$ and $\tau_\text{ep-DOS}$ with red and black dashed lines in Fig.~\ref{fig:bulkSi_resistivity}(b), respectively. The $\tau_\text{ep-DOS}$ resistivity follows the predicted $\tau_\text{ep-EPW}$, with the values slightly higher in the log scale. This result indicates that it may be reasonable to use the DOS approximation ($\tau_\text{ep-DOS}$) to predict the trends of resistivities at low carrier concentrations. This observation matches with our conclusion from a previous study as well~\cite{proshchenko2019optimization}. The inverted black triangles in Fig.~\ref{fig:bulkSi_resistivity}(b) represent measured bulk Si resistivities reported in Ref.~\citenum{mousty1974relationship}. The resistivity obtained with $\tau_\text{ep-EPW}$, closely matches with the measured data~\cite{mousty1974relationship} for n$_\text{e}<10^{17}$ cm$^{-3}$. At high doping, for n$_\text{e}$\textgreater$10^{17}$cm$^{-3}$, both the $\tau_\text{ep-EPW}$ and $\tau_\text{ep-DOS}$ results are lower than the experimental data. The consideration of IMS processes significantly influences the resistivity predictions at higher carrier concentrations and improves the match with the measured data. The solid red and black lines represent $\rho$ computed considering electron relaxation due to EPS-EPW+IMS and EPS-DOS+IMS, respectively. The increase of carrier concentration leads to higher electron scattering rate due to increased electron-impurity scattering. The inclusion of IMS processes does not influence the predictions at lower concentrations, as expected. The CRTA predicted $\rho$ (green line) closely matches the EPW+IMS predicted $\rho$ at lower concentrations, however, is higher than the EPW+IMS predictions at higher concentrations, highlighting the strong effect of different scattering mechanisms on $\rho$, especially in the high doping regime. The results shown in Fig.~\ref{fig:bulkSi_resistivity} establishes that it is essential to consider the electron relaxation times due to both EPS and IMS processes for accurate prediction of the Seebeck coefficient (diffusive contribution) and the resistivity of bulk Si at 300 K, especially at the high doping regime.
 
\section{Results and Discussion}

We now turn to the main focus of this study, which is to demonstrate the effect of the scattering processes on the electronic transport properties of Si/Ge SLs, with diverse lattice strain environments. The strain environment in fabricated Si/Ge heterostructures is highly nonuniform: $\sim3-4\%$ strain values have been reported in Si/Ge nanowire heterostructures with compositionally abrupt interfaces, grown via the VLS process~\cite{wen2015strain}. The lattice strain environment in SLs is strongly dependent on growth substrates~\cite{proshchenko2021role} and layer compositions~\cite{proshchenko2019optimization}. We illustrated in our previous publications that the layer compositions and external substrate induced strains prompt non-uniform monolayer separations, and modulate the Seebeck coefficients of Si/Ge SLs, within the CRTA~\cite{proshchenko2019optimization,proshchenko2021role,settipalli2020theoretical,pimachev2021first}. Here, we highlight the effect of EPS and IMS processes on the Seebeck coefficients, resistivities, and power-factors of these Si/Ge SLs with diverse strain environment. We also discuss how these predictions compare with the DOS scattering approximation and CRTA (SI Fig.~9) results. We compare the transport properties with the corresponding bulk Si properties for a better appreciation of the modulation of electronic transport in the SLs. 

\subsection{Effect of EPS on Electronic Transport Properties}

In the SLs with varied compositions and substrate induced strain environments, the energy bands shift due to lattice strain~\cite{proshchenko2019optimization,settipalli2020theoretical,proshchenko2021role}. Here, we illustrate the effect of these band energy shifts on the EPS processes and the resulting cross-plane transport properties of the SLs. Alongside, we discuss the applicability of the DOS scattering approximation by comparing the computed transport properties with the EPW predictions.

\begin{figure*}[t]
\includegraphics[width=1.0\linewidth]{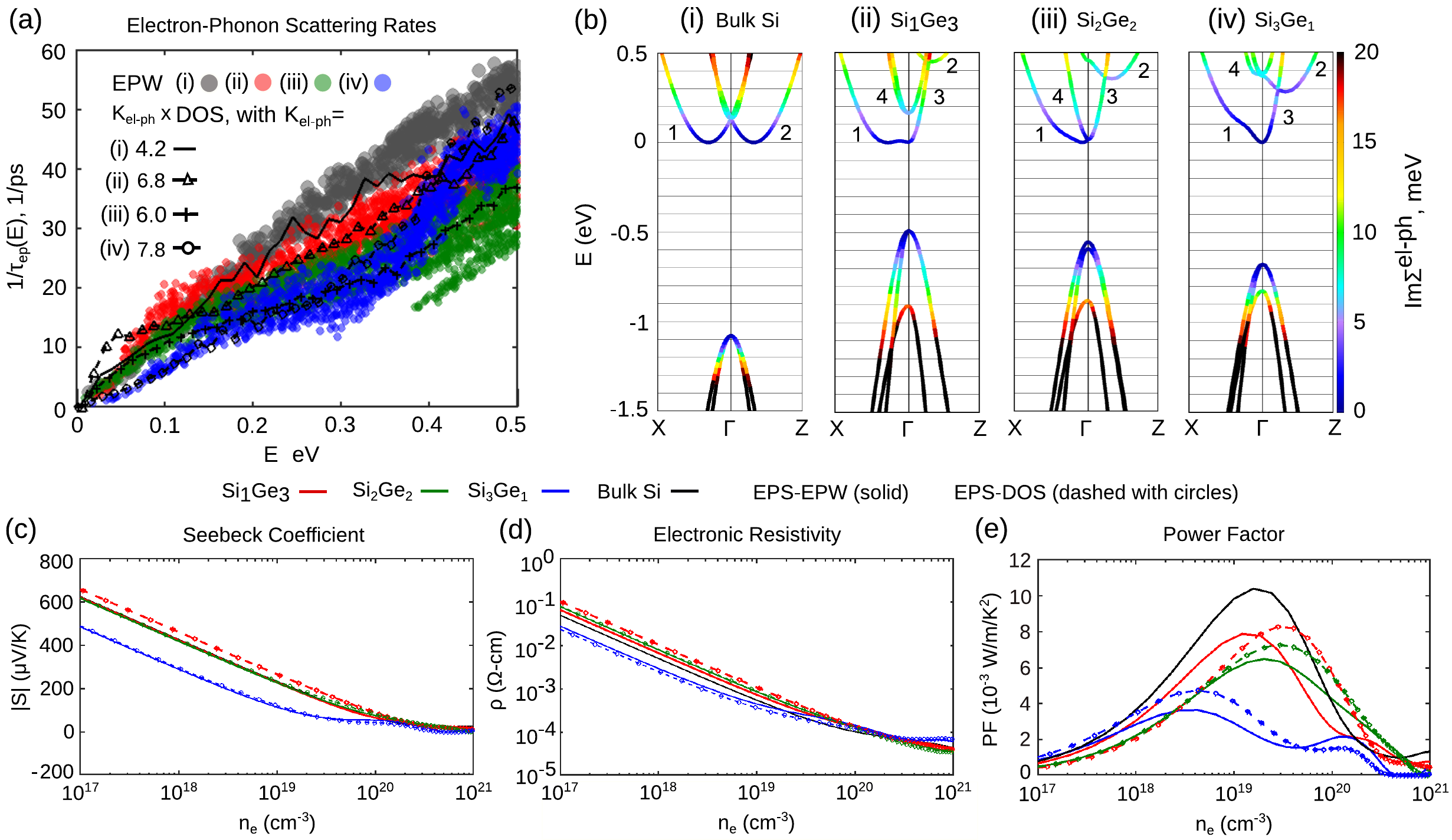}
\caption{Electronic properties of Si/Ge SLs with varied compositions: (ii) Si$_1$Ge$_3$, (iii) Si$_2$Ge$_2$, and (iv) Si$_3$Ge$_1$ SLs, grown on a fixed Si substrate. (a) Electron-phonon scattering rates predicted by the EPW approach (circles) and the DOS scattering approximation (dashed lines with symbols) ($1/\tau_\text{ep-DOS}(\text{E}) = \text{K}_\text{el-ph}\times\text{DOS}(\text{E})$). The proportionality constants, $\text{K}_\text{el-ph}$, are shown in respective legends. (b) Colormap of electronic band structures of (i) bulk Si and (ii) Si$_1$Ge$_3$, (iii) Si$_2$Ge$_2$, and (iv) Si$_3$Ge$_1$ SLs, weighted by the imaginary part of electron self energy. (c) Seebeck coefficients, (d) electronic resistivities and (e) power-factors of Si/Ge SLs incorporating EPS-EPW (solid) and EPS-DOS rates (dashed lines with circles).}
\label{fig:SLCompEPSOnly}
\end{figure*}

\subsubsection{Superlattices with Varied Compositions}

We investigate three SLs with different compositions, namely Si$_1$Ge$_3$, Si$_2$Ge$_2$, and Si$_3$Ge$_1$ SLs. The model SLs are assumed to be grown on a Si substrate. Figure~\ref{fig:SLConfigs} shows the representative configurations of the Si/Ge SLs investigated in the present study. We define the in-plane strain in the monolayers, $\Epsilon_{\text{Si}\parallel}$, with respect to bulk Si, as $\Epsilon_{\text{Si}\parallel}= (a_{\parallel}/a_\text{Si}-1)$. Here, $a_{\parallel}$ is the effective in-plane lattice constant of the SLs and $a_\text{Si}=5.40\;\text{\AA}$ is the LDA predicted bulk Si lattice constant. Using our definition, the in-plane lattice strain in the Si region of the SLs measures to $\Epsilon_{\text{Si}\parallel}=0 \%$. We show the EPW predicted EPS rates in (i) bulk Si (grey circles), (ii) Si$_1$Ge$_3$ (red circles), (iii) Si$_2$Ge$_2$ (green circles) and (iv) Si$_3$Ge$_1$ (blue circles), in Fig.~\ref{fig:SLCompEPSOnly}(a). We aligned the CBM of the different SLs to compare electron relaxation due to phonons in different systems. The E $=0$ eV point in the y-axis refers to the common conduction band minima (CBM) of the SLs. We show the EPS rates for $0\leq \text{E} \leq 0.5$ eV, because the Fermi levels of the three SLs, with n-type doping of $\sim 10^{17} - 10^{21}\ \text{cm}^{-3}$, are on the range of $\sim -0.14 - 0.45$ eV at 300 K. In order to establish a qualitative understanding of the EPS rates, we discuss the relationship between the imaginary part of the electron-phonon self energies, Im$\Sigma_{n\textbf{k}}$ and the SL band structures along symmetry directions. Although, the EPS rates are calculated from Im$\Sigma_{n\textbf{k}}$, by sampling the entire SL BZ, comparison of the electronic properties along SL symmetry directions reveals physical insights regarding electron relaxation in diverse SLs.

Figure~\ref{fig:SLCompEPSOnly}(b) shows the SL band structures in the X$-\Gamma-$Z symmetry directions, superimposed with Im$\Sigma_{n\textbf{k}}$. The panels in Fig.~\ref{fig:SLCompEPSOnly}(b) show the band structures of (i) bulk Si and (ii) Si$_1$Ge$_3$, (iii) Si$_2$Ge$_2$ and (iv) Si$_3$Ge$_1$ SLs, respectively, along the in-plane X$-\Gamma$ and cross-plane $\Gamma-$Z SL symmetry directions. The bulk Si bands are contributed from $\Delta$ valleys (panel (i)). The SL bands, shown in panels ((ii)-(iv)), are formed due to the combined effect of zone folding of bulk bands (panel (i)) dictated by structural symmetry, and periodic potential perturbation~\cite{proshchenko2021role,satpathy1988electronic,van1986theoretical,hybertsen1987theory,tserbak1993unified}. The varied compositions of these SLs result in diverse lattice strain environment, that regulates contributions from the valleys and modifies the SL bands~\cite{proshchenko2019optimization,settipalli2020theoretical}. As we increase the number of Si monolayers in the SLs, $L_{\text{Si}}= 1\rightarrow 3$ (Si$_1$Ge$_3$ $\rightarrow$ Si$_2$Ge$_2$ $\rightarrow$ Si$_3$Ge$_1$), for a fixed $L \ (L=4)$ and fixed $a_{\parallel} = a_{\text{Si}}$, the average cross-plane strain ($\propto (1- L_{Si}/L$)) decreases. The cross-plane strain is high in (ii) Si$_1$Ge$_3$ SL and decreases for (iii) Si$_2$Ge$_2$ and (iv) Si$_3$Ge$_1$ SLs. The reduced cross-plane strain induces a shift of type 2 conduction minibands, contributed by $\Delta_\perp$ valleys, to low energies in the $\Gamma-$Z direction. The minima of the type 2 $\Gamma-$Z conduction bands shift from (ii) 0.45 eV to (iii) 0.35 eV to (iv) 0.28 eV. The interplay between strain environment and confinement is more complex for the low energy type 1 and type 4 conduction bands in the $\Gamma-$X direction, formed by $\Delta_\parallel$ valleys. The intersections of type 1 bands with the $\Gamma$ point remain mostly same, however, the dispersive natures are significantly different. The colors in the band structures plot (Fig.~\ref{fig:SLCompEPSOnly}(b)) represent the corresponding imaginary part of the self-energies, Im$\Sigma_{n\textbf{k}}$, with values as indicated by the color bar at the side panel. The self-energies are small in the entire low energy conduction zone of interest E $=0-0.5$ eV, due to the absence of states in the bandgap region and also, low DOS in this energy range. However, color variations of the SL bands between panels ((ii)(green)$\rightarrow$(iv)(blue)) can be noted, when compared to the bulk bands of panel (i), that indicates change of the self-energies. The electron self-energies are directly influenced by the energy shifts and the dispersive nature of the SL bands, as we illustrate in the following discussion. The decrease of Im$\Sigma_{n\textbf{k}}$ of type 2 SL bands ((ii) \textgreater (iii) \textgreater (iv)) is due to the energy shift of the type 2 conduction minibands in the $\Gamma-$Z direction, induced by the changes of the cross-plane lattice strain environment. The energy shift reduces the availability of lower energy states and thus, lowers the probabilities of phonon-assisted transitions, resulting in smaller electron self-energies (green $\rightarrow$ blue). Similarly, the changes of self-energies associated with the type 4 bands ((ii), (iv)\textgreater (iii)), are connected to the nature of type 1 bands. The Im$\Sigma_{n\textbf{k}}$ associated with the type 1 bands remain small in all three SLs due to the absence of available lower energy states. However, the dispersive nature of these bands changes significantly in different SLs. The altered type 1 band induces changes of the electron self-energies in other SL bands, as shown in the band structures of the three panels ((ii)-(iv)). The increase of available density of states increases probabilities for phonon-assisted transitions, resulting in higher Im$\Sigma_{n\textbf{k}}$ for type 4 bands in panels (ii) and (iv). 

We illustrate below that the EPS rates are strongly connected to the nature of the SL bands and the self-energies, by highlighting the relationship between the resutls shown in panels Fig.~\ref{fig:SLCompEPSOnly}(a) and Fig.~\ref{fig:SLCompEPSOnly}(b). The EPS rates in the three SLs are shown in Fig.~\ref{fig:SLCompEPSOnly}(a). The EPS rates in all three SLs (red, green, blue) are lower than bulk Si (grey) for E $\gtrsim$ 0.2 eV due to low Im$\Sigma_{n\textbf{k}}$. At this energy range, Im$\Sigma_{n\textbf{k}}$ values in bulk Si are on the range $\sim$ 15-20 meV (red-black), higher than the values in the SLs (green-yellow). Comparing the EPS rates in (ii) Si$_1$Ge$_3$ SLs (red) and (iii) Si$_2$Ge$_2$ SLs (green), one can note that they match till E \textless 0.1 eV. Above this energy range, the rates in (ii) Si$_1$Ge$_3$ SLs (red) increase at a faster rate. This increase could be attributed to the higher self-energies of type 4 bands of (ii) Si$_1$Ge$_3$ SLs, compared to those in (iii) Si$_2$Ge$_2$ SLs, in this energy range. As we discussed previously, the higher self-energy of type 4 bands of (ii) Si$_1$Ge$_3$ SLs is influenced by the nature of type 1 bands. The lowest EPS rates are observed for (iv) Si$_3$Ge$_1$ SL at most energies (blue circles) due to the small electron self-energies of (iv) Si$_3$Ge$_1$ SL bands. However, the results match the rates in (ii) Si$_1$Ge$_3$ SLs (red) at E $\gtrsim$ 0.4 eV, due to the increase of self-energies of electrons, especially in type 4 bands (Fig.~\ref{fig:SLCompEPSOnly}(b)-(iv)). At E $\gtrsim$ 0.4 eV, the electrons in type 4 bands also participate in the scattering of process. The above discussion illustrates that the EPS rates are controlled by the nature of the SL bands. Consequently, the DOS scattering approximation could be used to describe the electron relaxation processes reasonably well. The EPS-DOS rates ($1/\tau_\text{ep-DOS}$), are shown in Fig.~\ref{fig:SLCompEPSOnly}(a) with (i) black solid line (bulk Si) and black dashed lines with (ii) triangles (Si$_1$Ge$_3$), (iii) +'s (Si$_2$Ge$_2$) and (iv) circles (Si$_3$Ge$_1$), respectively. As can be seen from Fig.~\ref{fig:SLCompEPSOnly}(a), the EPS-DOS rates follow the EPW predictions closely. The proportionality constants, $\text{K}_\text{el-ph}$, are given by (ii) 6.8, (iii) 6.0 and (iv) 7.8, respectively. The slight variations of $\text{K}_\text{el-ph}$ are connected to the varying DOS in the three SLs. The electronic DOS is low in (iv) Si$_3$Ge$_1$ compared to the other SLs for the energy range of interest. However, the EPS rates in the three SLs are mostly similar. The lower DOS and similar EPS rates result in higher K$_{\text{el-ph}}$ for (iv) Si$_3$Ge$_1$ ($1/\tau_{ep}(\text{E}) = \text{K}_\text{el-ph}\uparrow\times\text{DOS}(\text{E})\downarrow$). We like to point out that the determination of the scaling constants , K$_{\text{el-ph}}$, is highly dependent on the fitting procedure. As we show in Fig.~4 of the SI document, different values of the scaling constants  could be obtained for relaxed Si$_2$Ge$_2$ SL, depending on the numerical procedure. The variations of K$_{\text{el-ph}}$ do not change the resulting S but affect $\rho$ (SI Fig.~5). This discussion highlights the importance of exercising caution while using the DOS scattering rates to compute electronic transport properties. 

\begin{figure*}[ht]
\includegraphics[width=1.0\linewidth]{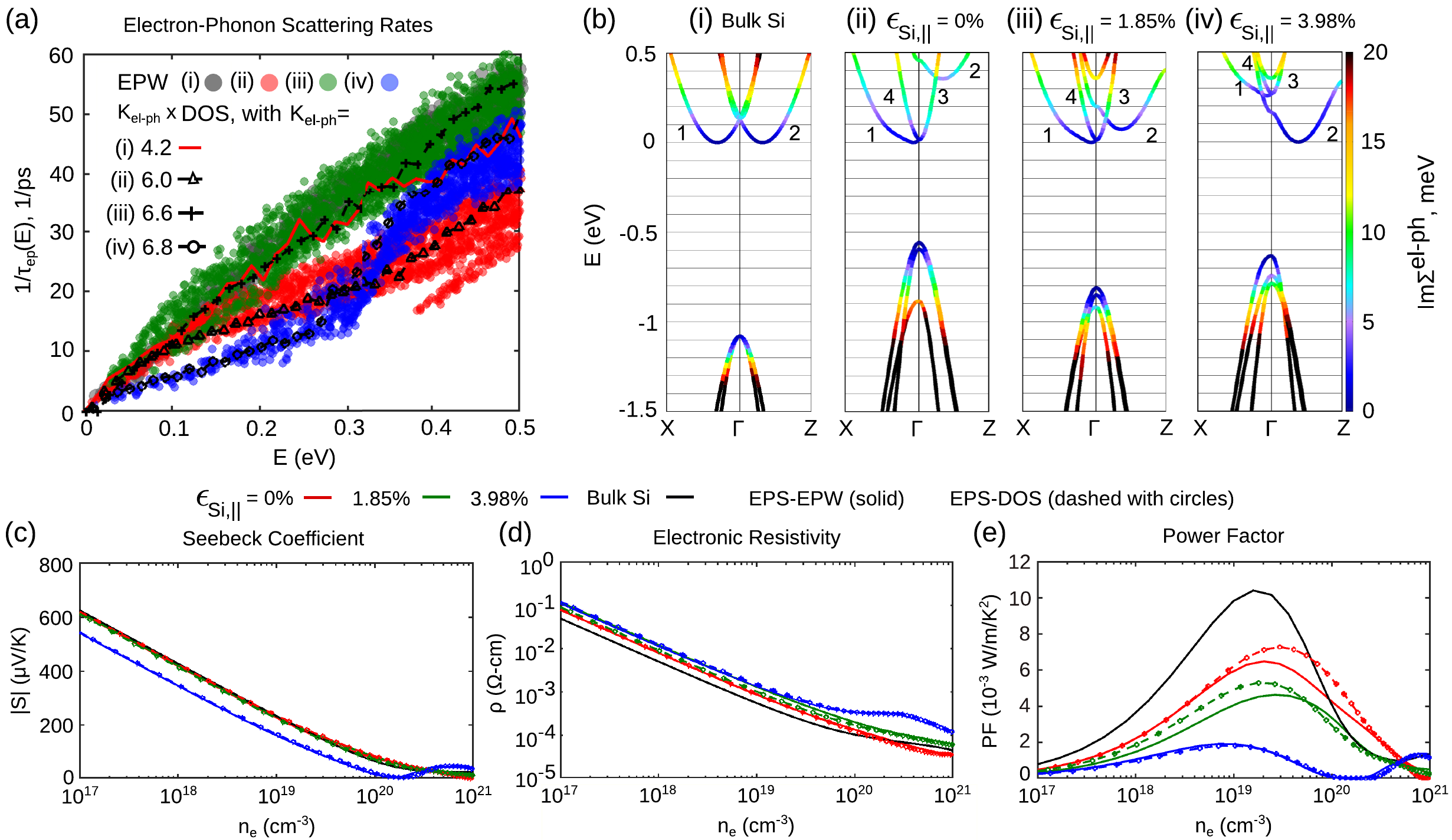}
\caption{Electronic transport properties of (i) bulk Si and Si$_2$Ge$_2$ SLs with in-plane strain (ii) 0\%, (iii) 1.85\%, and (iv) 3.98\%, induced by growth substrates. The proportionality constants, $\text{K}_\text{el-ph}$, are shown in respective legends. (a) Electron-phonon scattering rates predicted by the EPW approach (circles) and the DOS scattering approximation (dashed lines with symbols). (b) Colormap of electronic band structures of (i) bulk Si and Si$_2$Ge$_2$ SLs with (i) 0\%, (ii) 1.85\%, and (iii) 3.98\% in-plane strains, weighted by the imaginary part of electron self energy. (c) Seebeck coefficients, (d) electronic resistivities and (e) power-factors of strained Si/Ge SLs incorporating EPS-EPW (solid) and EPS-DOS rates (dashed lines with circles).}
\label{fig:SLStrainEPSOnly}
\end{figure*}

The electronic transport coefficients: Seebeck coefficients (Eq.~\ref{eq:seebeck}), resistivities (Eq.~\ref{eq:sigma}), and power-factors are obtained from the $\mathcal{L}$-integrals (Eq.~\ref{eq:Lintegral}). The strain-modulated DOS, electron group velocities and electron scattering rates affect the energy-dependence of the TDFs, as depicted in Eq.~\ref{eq:TDF}. The overlap of the energy dependent TDFs with the Fermi window functions $\left((\text{E}-\text{E}_\text{F})^{n}\left(-\frac{\partial f}{\partial \text{E}} \right)\right)$ results in modulated $\mathcal{L}$-integrals, and the electronic transport coefficients. In Fig.~\ref{fig:SLCompEPSOnly}(c) and (d), we show the S and $\rho$ of Si$_1$Ge$_3$ (red), Si$_2$Ge$_2$ (green), and Si$_3$Ge$_1$ SLs (blue), respectively, computed incorporating $\tau_\text{ep-EPW}$ (solid lines) and $\tau_\text{ep-DOS}$ (dashed lines with circles). The S of the Si$_1$Ge$_3$ and the Si$_2$Ge$_2$ SLs follow the bulk Si values closely. The similar Seebeck coefficients of (ii) and (iii) SLs can be explained by noting that these two systems share similar DOS, resulting in similar K$_{\text{el-ph}}$ values. The DOS in (iv) Si$_3$Ge$_1$ SLs is somewhat lower, resulting in higher K$_{\text{el-ph}}$. The S of Si$_3$Ge$_1$ deviates from the bulk Si values and displays reduced values at most carrier concentrations, n$_\text{e}$. We note that S of Si$_3$Ge$_1$ SL shows a non-monotonic behavior with increase of n$_\text{e}$. The Seebeck value is low compared to all other cases for n$_\text{e} \lesssim 10^{20}$cm$^{-3}$ ($\text{E}_\text{F} \sim$ 2.5 eV) and matches them at high n$_\text{e}$. The behavior can be explained by discussing the band structure shown in Fig.~\ref{fig:SLCompEPSOnly}(b)(iv). The lattice strain in Si$_3$Ge$_1$ SL increases the energy separation between the type 1 and type 4 $\Delta_\parallel$ bands. The gap between energy levels decreases the overlap between TDF and the Fermi window, causing S to reduce at n$_\text{e}<2\times10^{20}$cm$^{-3}$ ($\text{E}_\text{F}$ \textless 0.25 eV). However, when the Fermi level ($\text{E}_\text{F}$) is aligned close to the edge of the $\Delta_\perp$ minibands (type 2), S increases at high doping regime.

The effect of varied scattering rates or inversely the electron relaxation times is more pronounced on the resistivities of the SLs. The low EPS rates in Si$_3$Ge$_1$ SLs indicate high relaxation times and consequentially, low resistivities. The $\rho$ of Si$_3$Ge$_1$ SLs is lower than the bulk values at most carrier concentrations, $\text{n}_\text{e}\lesssim\times10^{20}$ cm$^{-3}$. The $\rho$ of Si$_1$Ge$_3$ and Si$_2$Ge$_2$ SLs remain close to the bulk values due to similar EPS rates. The resistivities obtained with EPS-DOS rates deviate slightly from the EPW predictions. The difference between the electronic transport property predictions from the two approaches become more apparent when S and $\rho$ are combined together to calculate PF ($\text{S}^2/\rho$). The variations of S and $\rho$ lead to considerable modulations of PF, especially at high $\text{n}_\text{e}$. The PF of Si$_1$Ge$_3$ SL is $\sim1.4$-fold larger than bulk Si at $\text{n}_\text{e}\sim2.7\times10^{20}$ cm$^{-3}$, while the PF of Si$_3$Ge$_1$ PF remains lower than that of bulk Si for all $\text{n}_\text{e}$. The lowest PF ($\text{S}^2/\rho$) of Si$_3$Ge$_1$ SL could be attributed to the strong reduction of S at most carrier concentrations. Overall, we note that the peak PFs of the chosen Si/Ge SLs are not higher than that of bulk Si. We like to point out that most studies of thermoelectric properties of phonon engineered systems assume that electronic properties remain unaffected. Here, we show that compositions of SLs modulate the SL bands and their EPS processes. The modulations lead to strong reduction of PFs at most n$_\text{e}$ compared to bulk Si. We like to point out that one needs to exercise caution while performing the numerical evaluation of the $\mathcal{L}$-integrals. In Si Fig.~6 and Fig.~7, we show how different alignments between EPS rates and DOS near the CB edge can affect the integrals and impact both S and $\rho$ (SI Fig.~7). 

\subsubsection{Superlattices with Lattice Strain Induced by Substrates}

To model SLs with substrate induced strain environments, we consider Si$_2$Ge$_2$ SLs grown on Si, Si$_{0.5}$Ge$_{0.5}$, and Ge substrates, respectively. Due to the lattice mismatch, the substrates induce in-plane strains of $\Epsilon_{\text{Si}\parallel}=$ $0 \%$, $1.85 \%$, and $3.98 \%$, in the Si region of the SL. In Fig.~\ref{fig:SLStrainEPSOnly}(a), we show the EPS rates in substrate strained Si$_2$Ge$_2$ SLs, computed with two different approaches. The EPS-EPW rates, $1/\tau_\text{ep-EPW}$, are shown with (i) grey (bulk Si), (ii) red (0\%), (iii) green (1.85\%) and (iv) blue (3.98\%) circles. The EPS-DOS rates are shown with (i) red solid line (bulk Si) and black dashed lines with (ii) triangles (0\%), (iii) +'s (1.85\%) and (iv) circles (3.98\%), respectively. Similar to the previous case, E $=0$ eV point in the y-axis refers to the common CBM of the strained SLs. We choose a similar x-axis range because the Fermi levels of the three strained SLs with n-type doping of $\sim 10^{17} - 10^{21}\ \text{cm}^{-3}$ are on the range $\sim -0.13 - 0.44$ at 300 K. In order to develop a qualitative understanding of the trends of the EPS rates in the strained SLs, we discuss the variations of the imaginary part of the electron-phonon self energies, Im$\Sigma_{n\textbf{k}}$, in relation to the SL band structures along different symmetry directions. As we have demonstrated for the SLs with varied compositions grown on fixed substrates, the comparison of the electronic properties along different symmetry directions provides physical understanding regarding the EPS rates in SLs.

Figure~\ref{fig:SLStrainEPSOnly}(b) shows the SL band structures along the in-plane X$-\Gamma$ and cross-plane $\Gamma-$Z symmetry directions, superimposed with Im$\Sigma_{n\textbf{k}}$. The panels in Fig.~\ref{fig:SLStrainEPSOnly}(b) shows the band structures of (i) bulk Si and Si$_2$Ge$_2$ SLs with in-plane strain of (ii) 0\%, (iii) 1.85\% and (iv) 3.98\%, respectively. The SL bands shown in panels ((ii)-(iv)) are formed due to the combined effect of zone folding of bulk bands, and periodic potential perturbation. The substrate induced lattice strain induces valley splittings and regulates contributions from the split valleys to modify the SL bands~\cite{proshchenko2021role,settipalli2020theoretical}. With the increase of $\Epsilon_{\text{Si}\parallel}$ ((ii)$\rightarrow$(iv)), the low energy conduction bands (type 1, type 4) in the $\Gamma-$X direction, formed by the $\Delta_\parallel$ valleys, move upward in energy. The strain-induced band shifts can be particularly noted by considering the upward shift of the conduction bands and downward shift of valence bands together. It is good to keep in mind that we aligned the CBM of the different SLs to E $=0$ eV in these plots. The intersection of the lowest $\Gamma-$X conduction band (type 1), with the $\Gamma$ point shifts upward from (ii) 0.01, (iii) 0.01 to (iv) 0.27. On the other hand, the low energy conduction minibands (type 2) in the $\Gamma-$Z direction, contributed by the $\Delta_\perp$ valleys, shift downward with the increase of in-plane strain. The intersection of these bands with the $\Gamma$ point moves from (ii) 0.46 to (iii) 0.20 (iv) to 0.17 eV. The other set of parabolic conduction bands (type 3) along $\Gamma-$Z direction, formed due to the overlap of $\Delta_\parallel$ valleys at the $\Gamma$ point, shifts upward in energy together with the type 1 bands. It can be noted that the CBM shifts from the $\Delta_\parallel$ bands (type 1) to the $\Delta_\perp$ bands (type 2), along cross-plane $\Gamma-Z$ direction, with the increase of $\epsilon_{\text{Si}\parallel}$. Similar energy shifts of the strain-split $\Delta$ valley bands have been observed in other short-period SLs (e.g., Si$_4$Ge$_4$)~\cite{proshchenko2021role}.

The colors of the band structures plot (Fig.~\ref{fig:SLStrainEPSOnly}(b)) represent the corresponding imaginary part of the self-energies. The self-energy values are as indicated by the color bar at the side panel. The self-energies are small in the entire low energy conduction zone of interest $0-0.5$ eV, due to the absence of states in the bandgap region. The low self-energies are also due to the small density of CB states in this region, however, the density varies for different strain cases leading to variations of the self-energies. We illustrate in the following that the changes of the self-energies are related to the strain induced energy shifts of the SL bands. As discussed above, the energy of the type 2 $\Delta_\perp$ valley band lowers ((ii) $\rightarrow$ (iv)) with the increase of in-plane strain, along the cross-plane $\Gamma-Z$ direction. The self-energy of the shifted $\Delta_\perp$ band decreases with increase of strain (blue-green (ii) $\rightarrow$ blue (iv)), due to the reduced availability of lower energy states. Additionally, the energy gap between the type 2 band and other bands (3, 4) increases due to the energy shift, as can be noted from panel (iv) of Fig.~\ref{fig:SLStrainEPSOnly}(b). The increase of the energy gap lowers the probabilities of phonon-assisted energy transitions. As a result, the self-energies remain low till E $\lesssim$ 0.3 eV. For E \textgreater 0.3 eV, the self-energy of the bands of the (iv) 3.98\% strained SL increases due to increased probabilities for phonon-assisted transitions. We show the colormap of band structures and Im$\Sigma_{n\textbf{k}}$ of the region $-2$ to $2$ eV, in SI Fig.~1, for a broader perspective. 

These variations of the self-energies with E directly influence the EPS rates in 3.98\% strained Si$_2$Ge$_2$ SL, as can be seen from Fig.~\ref{fig:SLStrainEPSOnly}(a) (blue circles). The EPS rates are low till (E $\sim$ 0.3 eV) and increase sharply afterwards. Similarly, one can relate the EPS rates in the Si$_2$Ge$_2$ SLs with (ii) 0\% (red) and (iii) 1.85 \% (green) in-plane strain environment, to the nature of the bands of the two strained SLs. As can be seen from panels (ii) and (iii) of Fig.~\ref{fig:SLStrainEPSOnly}(b), the  band structures of type 1, 3, and 4 bands are quite similar near the CB edge. As a consequence, the EPS rates in these two SLs match for E $\lesssim$ 0.1 eV. For E $\gtrsim$ 0.1 eV, the EPS rates for case (iii) (green) are higher than those for case (ii). The increased EPS rates for case (iii) SL are due to the following reasons. The scattering rates of electrons in the type 2 $\Delta_\perp$ valley band contribute to the EPS rates of case (iii) SL at approximately, E $\sim$ 0.1 eV. The type 2 band of the (iii) 1.85\% strained SL is shifted to lower energy due to increased strain. Additionally, the small energy gap between the bands increases probabilities for phonon-assisted transitions, resulting in higher EPS rates for case (iii) (green). The EPS rates for case (ii) SL (red) increases around E $\gtrsim 0.4$ eV once electrons in type 2 band participate in the scattering process. The DOS scattering rates follow the EPW predictions closely, with proportionality constants, $\text{K}_\text{el-ph}$, given by (ii) 6.0, (iii) 6.6 and (iv) 6.8, respectively. Overall, we note from Fig.~\ref{fig:SLCompEPSOnly}(a) and Fig.~\ref{fig:SLStrainEPSOnly}(a) that the DOS scattering approximation can describe the trends of the EPS rates of short-period superlattices. However, it is essential to determine $\text{K}_\text{el-ph}$ accurately to predict electronic transport properties. $\text{K}_\text{el-ph}$ is determined by aligning DOS data with the EPS-EPW rates. As we have discussed previously, $\text{K}_\text{el-ph}$ is highly dependent on the numerical fitting procedure (SI Fig.~4-5). Therefore, it is difficult to solely rely on this approach to predict the transport properties without the knowledge of higher accuracy EPS rates (Fig.~\ref{fig:SLCompEPSOnly}(e) and Fig.~\ref{fig:SLStrainEPSOnly}(e)). 

Figure~\ref{fig:SLStrainEPSOnly}(c) and (d) show the S and $\rho$ of Si$_2$Ge$_2$ SLs with in-plane strain of $\Epsilon_{\text{Si}\parallel}=$ $0 \%$ (red), $1.85 \%$ (green), and $3.98 \%$ (blue), respectively, computed incorporating $\tau_\text{ep-EPW}$ (solid lines) and $\tau_\text{ep-DOS}$ (dashed lines with circles). We note that S of the 3.98\% strained Si$_2$Ge$_2$ SL (blue) shows a non-monotonic behavior with increase of n$_\text{e}$. The Seebeck value peaks at high n$_\text{e}$ preceded by values lower than those of bulk Si. Such results are uncommon since the Seebeck coefficients of doped semiconductors and metals are expected to decrease monotonically with increase of carrier concentration, in agreement with the Pisarenko relation. The S modulations are directly related to the nature of bands in the strained SLs, as can be illustrated by discussing the band structure shown in panels of Fig.~\ref{fig:SLStrainEPSOnly}(b). Comparing the band structures shown in panels (iii) and (iv), it can be observed that the energy gap between type 2 ($\Delta_\perp$) and type 1, 3, 4  ($\Delta_\parallel$) bands increases due to the substrate-strain induced shift of SL energy bands. The gap between energy levels results in low DOS near CB edge and decreased overlap between TDF and the Fermi window (Eq.~\ref{eq:Lintegral}). As a result, S is reduced at n$_\text{e}<2.6\times10^{20}$cm$^{-3}$ ($\text{E}_\text{F}$ \textless 0.25 eV) compared to that of bulk Si. However, when the Fermi level ($\text{E}_\text{F}$) is aligned close to the edge of the $\Delta_\parallel$ minibands (type 1, 3, 4), the higher DOS leads to a S-peak at high $\text{n}_\text{e}$. Such substrate strain modulated non-monotonic Seebeck coefficients have been observed in other short-period SLs~\cite{proshchenko2021role,bahk2012seebeck,proshchenko2019optimization,settipalli2020theoretical,pimachev2021first}. 

The $\rho$ of Si$_2$Ge$_2$ SLs with 3.98\% in-plane strain also displays nonmonotonic behavior, due to the nature of the $\Delta_\parallel$ and $\Delta_\perp$ bands in the strained SL (Fig.~\ref{fig:SLStrainEPSOnly}(d)). The resistivities of the SLs are greater compared to that of bulk Si at most n$_\text{e}$. $\rho$ of 3.98\% strained SL (iv, blue) are higher than those of (ii) (red) and (iii) (green) SLs at all n$_\text{e}$ of interest, with a noticeable increase at n$_\text{e}\sim2.2\times10^{20}$cm$^{-3}$ ($\text{E}_\text{F}$ $\sim$ 0.25 eV). The cross-plane electrical resistivity $\rho$ can be expressed as, $\rho\propto1/\int\tau(\text{E})v_z^2(\text{E})\text{DOS}(\text{E})\left (\partial f/d\text{E}\right ) d\text{E}$ (Eq.~\ref{eq:Lintegral}-\ref{eq:sigma}). As illustrated previously, the DOS scattering approximation ($1/\tau_\text{ep-DOS}(\text{E}) = \text{K}_\text{el-ph}\times\text{DOS}(\text{E})$) could describe the EPS rates in strained Si$_2$Ge$_2$ SLs reasonably well. As shown in Fig.~\ref{fig:SLCompEPSOnly} and Fig.~\ref{fig:SLStrainEPSOnly}, S and $\rho$ predicted using $\tau_\text{ep-DOS}$(E) (dashed lines with circles) follow the $\tau_\text{ep-EPW}$(E) (solid lines) predictions closely, for carefully chosen values of $\text{K}_\text{el-ph}$. Inserting this approximation, $\rho$ can be simplified to, $\rho\propto1/\int v_z^2(\text{E})\left (\partial f/d\text{E}\right ) d\text{E}$. This expression indicates that electron group velocities play a significant role in determining $\rho$. We show a comparison between $v_z^2(\text{E})$ in the strained Si$_2$Ge$_2$ SLs in SI Fig.~8. The results show that $v_z^2(\text{E})$ in the 3.98\% strained SL, lies below those of the other two SLs for all energies of interest, resulting in high resistivities. Between $0\leq E \lesssim 0.14$ eV, $v_z^2(\text{E})$ increases fast with E, causing decrease of $\rho$ with n$_\text{e}$ up to n$_\text{e}\sim9.5\times10^{19}$cm$^{-3}$. For 0.14 \textless E \textless 0.3 eV, $v_z^2(\text{E})$ remains comparatively `flat' causing the `bump' in the resistivity. $v_z^2(\text{E})$ increases again for E \textgreater 0.3 eV. The product of Fermi window and $v_z^2(\text{E})$, $v_z^2(\text{E})\left (\partial f/d\text{E}\right )$, encompasses the fast increase of $v_z^2(\text{E})$ and results in the change of trend, followed by a steady decrease of $\rho$. The resistivity trend starts to change near E $\sim 0.22$ eV (n$_\text{e}\sim1.97\times10^{20}$cm$^{-3}$).

\begin{figure*}[ht]
\includegraphics[width=1.0\linewidth]{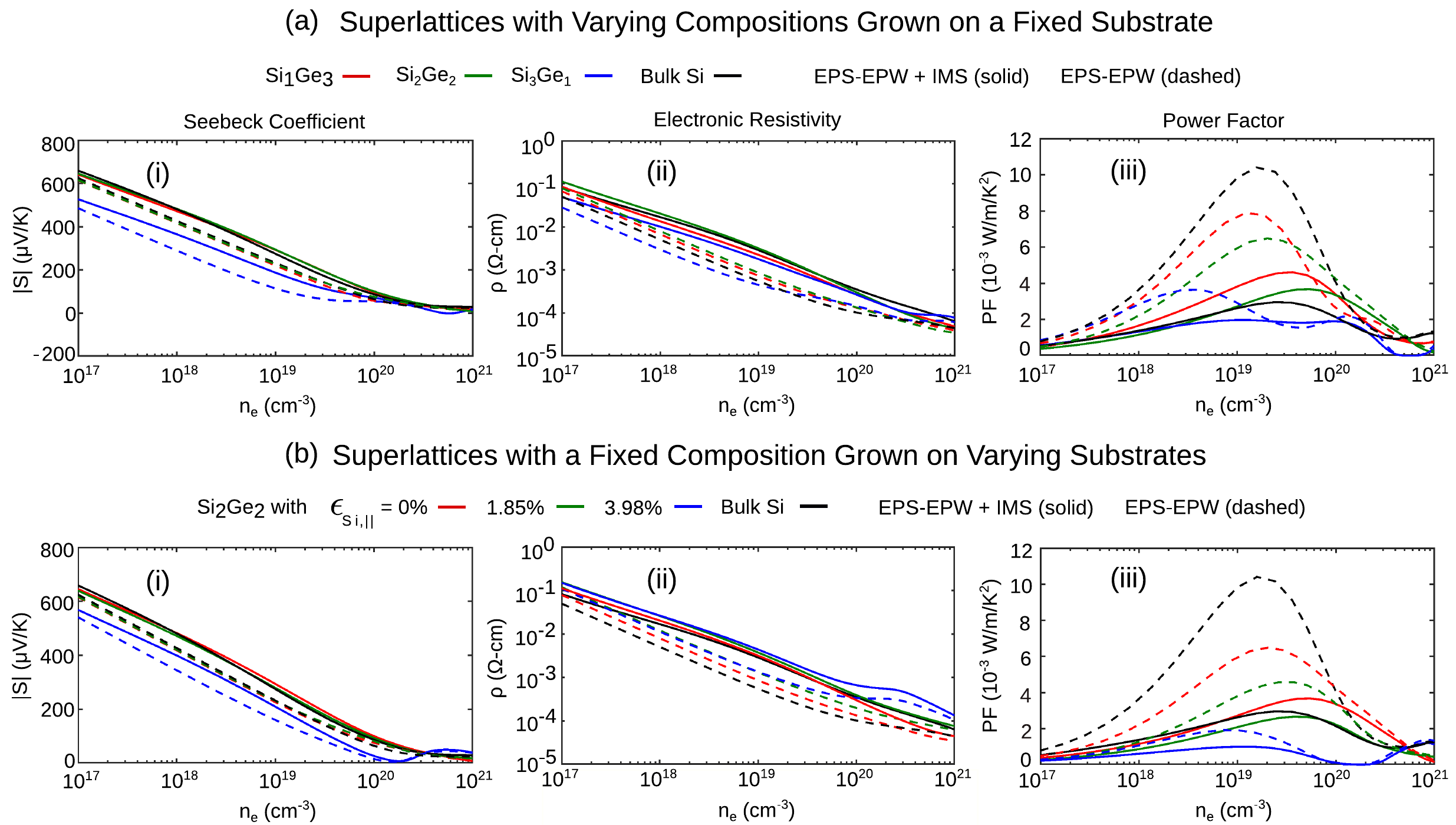}
\caption{(i) Seebeck coefficients, (ii) electronic resistivities and (iii) power-factors of SLs with diverse strain environments: (a) Si/Ge SLs with varied compositions, grown on identical substrates, (b) Si$_2$Ge$_2$ SLs grown on different substrates. Solid lines represent predictions made by considering electron scattering due to phonons and ionized impurities. Dashed lines display results obtained by considering electron scattering due to phonons only, for comparison.}
\label{fig:SLCompStrainEPSIMS}
\end{figure*}

The non-monotonic S and $\rho$ lead to significant modulations in the power-factors, as shown in Fig.~\ref{fig:SLStrainEPSOnly}(e). S and $\rho$ predicted using $\tau_\text{ep-DOS}$(E) (dashed lines with circles) match the $\tau_\text{ep-EPW}$(E) (solid lines) predictions, however, the two PF predictions show considerable differences. We note that the PFs for all strained Si$_2$Ge$_2$ SLs are low compared to bulk Si, for n$_\text{e}<9.1\times10^{19}$cm$^{-3}$, with the values decreasing with increase of substrate induced strain. However, some of the strained SLs show an increased PF at high doping regime, n$_\text{e}$ \textgreater $10^{20}$cm$^{-3}$. This follows from the band structure physics of Si/Ge SLs grown on substrates inducing high strains. The strain induces shift of $\Delta_\parallel$ valley minibands to higher energies, resulting in closely spaced minibands at high n$_\text{e}$ or high $\text{E}_\text{F}$. The dense miniband structures, the resultant electronic group velocities and scattering rates directly result in a$~\sim$1.8 fold improvement of PF over bulk Si at n$_\text{e}\sim2.7\times10^{20}$cm$^{-3}$, for Si$_2$Ge$_2$ SL with $\Epsilon_{\text{Si}\parallel}=0 \%$ (red, grown on Si substrate). However, the peak PFs of the SLs are not improved compared to bulk Si at most carrier concentrations, to prove beneficial for thermoelectric applications. 

\subsection{Effect of IMS on Electronic Transport Properties}

In addition to electron scattering due to phonons, electronic transport properties of heavily doped semiconductor superlattices are strongly influenced by scattering due to impurities, which has hardly been investigated using first-principles approaches. As outlined in the Methods section, we obtain the electron relaxation times due to the IMS processes, $\tau_\text{im}$ and combine with those due to the EPS processes, $\tau_\text{ep-EPW}$ using Matthiessen's rule, to calculate the total $\tau$(E). In Fig.~\ref{fig:SLCompStrainEPSIMS}, we show the electronic transport properties of (a) SLs of varied compositions, grown on Si substrates, and (b) Si$_2$Ge$_2$ SLs grown on different substrates, respectively. The solid lines represent predictions obtained by considering the total energy-dependent electron relaxation time, $\tau$(E). The dashed lines represent predictions obtained by considering the electron-phonon relaxation time, $\tau_\text{ep-EPW}$(E), only. The dashed lines display the same results we have shown in Fig.~\ref{fig:SLCompEPSOnly} and \ref{fig:SLStrainEPSOnly}, obtained using the EPS-EPW rates. We include both the predictions in the same figure to compare the effect of different scattering mechanisms on the electronic transport properties of the SLs. 

Figure~\ref{fig:SLCompStrainEPSIMS}(a-i) show the Seebeck coefficients of SLs of varied compositions, grown on Si substrates. The comparison between the solid lines and the dashed lines exhibit the change of S results, due to different scattering processes. This is unlike the results reported for bulk Si, that the diffusive part of S does not depend strongly on the nature of the scattering mechanisms~\cite{fiorentini2016thermoelectric}. A maximum 1.77-fold increase of S can be observed for Si$_{3}$Ge$_{1}$ SL (blue) when considering both EPS and IMS processes, compared to EPS only results. However, the trend of S for the different SLs remains preserved. The Seebeck coefficients of Si$_{1}$Ge$_{3}$ (red) and Si$_{2}$Ge$_{2}$ SLs (green) remain close to S of bulk Si (black) at most $\text{n}_\text{e}$, while S of Si$_{3}$Ge$_{1}$ SL (blue) is reduced at all $\text{n}_\text{e}$ (Fig.~\ref{fig:SLCompStrainEPSIMS}(a-i)). Figure~\ref{fig:SLCompStrainEPSIMS}(b-i) shows the resistivities of SLs obtained by considering electron scattering only due to phonons (dashed lines) compared to those obtained by considering both EPS and IMS processes (solid lines). Similar to S, the resistivities of Si$_{3}$Ge$_{1}$ SL (blue) are reduced, while Si$_{1}$Ge$_{3}$ (red) and Si$_{2}$Ge$_{2}$ SLs (green) values are closer to bulk Si values (black) at most n$_\text{e}$. The consideration of the IMS processes increases the resistivities at almost all carrier concentrations, as expected. The effect is more pronounced, as shown in the log scale. A maximum 4.3-fold increase of $\rho$ is observed for the Si$_3$Ge$_1$ SL (blue), when considering both EPS and IMS processes, compared to EPS only results. Figure~\ref{fig:SLCompStrainEPSIMS}(a-iii) shows the comparison between power-factors of SLs obtained by considering only EPS processes (dashed lines) and both EPS and IMS processes (solid lines). The consideration of IMS processes significantly increase resistivities and reduce PFs for all systems. However, the variations of S and $\rho$ lead to a peak PF of 4.60 for Si$_{1}$Ge$_{3}$ SL (red) at n$_\text{e} \sim 3.7\times10^{19}$cm$^{-3}$, which is $\sim1.56$-fold (or $56\%$) greater than the bulk Si peak PF of $2.95$ at n$_\text{e}\sim2.4\times10^{19}$cm$^{-3}$. The maximum increase of PF over bulk Si value can be noted for n$_\text{e}\sim7.4\times10^{19}$cm$^{-3}$. The PF of Si$_{1}$Ge$_{3}$ SL is $\sim1.62$-times the bulk Si PF value at this carrier concentration. 

We make similar observations regarding the role of IMS processes on the electronic transport properties of Si$_{2}$Ge$_{2}$ SLs, grown on substrates that induce different in-plane strains. The S results shown with solid lines in Fig.~\ref{fig:SLCompStrainEPSIMS}(b-i) are modified from the results shown with dashed lines, that only considers electron relaxation due to the EPS processes. A maximum 1.4-fold increase of EPS+IMS S results is observed for Si$_2$Ge$_2$ SL on Si substrate (red). Figure~\ref{fig:SLCompStrainEPSIMS}(b-ii), shows the resistivities obtained by considering EPS only (dashed lines) compared to those obtained by considering both EPS and IMS processes (solid lines). Similar to the SLs with varied compositions, we note that the consideration of the IMS processes increases the resistivities at almost all carrier concentrations. A maximum 3.7-fold increase of $\rho$ is observed for the Si$_2$Ge$_2$ SL on Si substrate (red). Consequentially, the PFs of these SLs, as shown with solid lines in Fig.~\ref{fig:SLCompStrainEPSIMS}(b-iii), are significantly different from the EPS only results (dashed lines). The comparison between the solid and the dashed lines shows that the incorporation of the combined $\tau$(E) results in reduced PFs compared to the EPS only results, for all systems. However, we find that the PF of the SL grown on Si substrate (red) is higher than that of bulk Si for n$_\text{e}\sim7.6\times10^{18}-5.2\times10^{20}$ cm$^{-3}$. A maximum $\sim1.9$-fold improvement is observed at n$_\text{e}\sim2.3\times10^{20}$cm$^{-3}$. Peak to peak comparisons between PFs of different systems show that peak PF of the SL grown on Si substrate is $3.67$ at n$_\text{e}\sim5.5\times10^{19}$cm$^{-3}$. This value is $\sim1.24$-fold (or $24\%$) greater than the bulk Si peak PF of $2.95$ at n$_\text{e}\sim2.4\times10^{19}$cm$^{-3}$.

Overall, we observe that the consideration of both the scattering processes decreases PFs of bulk Si and Si/Ge SLs, compared to the results, obtained by only including EPS processes. However, the PFs of the SLs are comparatively less affected than bulk Si, when the impurity scattering mechanisms are included. This observation can be explained through a qualitative discussion of electron mean free paths (MFPs) in bulk Si and Si/Ge heterostructures, and the effect of IMS processes on the MFPs. First-principles studies have reported that the electron MFPs can go up to about 60 nm in bulk silicon at 300 K at low carrier concentration, $10^{16}\ \text{cm}^{-3}$. The MFPs in Si strongly depend on carrier concentrations, and decrease monotonically with increase of carrier concentration, due to the electron-impurity scattering processes. The MFPs reduce to about 20 nm in bulk Si with carrier concentration of $10^{19}\ \text{cm}^{-3}$. In this doping regime, the dominant contribution to the electrical conductivity comes from electrons with MFPs less than 10 nm~\cite{qiu2015first}. Such studies about electron MFPs in Si/Ge heterostructures are limited, however, the profiles of the chosen short-period SLs introduce potential perturbations at length scales smaller than the electron MFPs contributing to electron transport~\cite{tsu1973tunneling}. Therefore, contribution from electrons with lower MFPs can be expected to dominate, which are not considerably affected by the IMS processes. This argument could be used to explain the observation that the influence of IMS processes on the electronic transport properties of short-period SLs is less severe compared to bulk Si.

\section{Summary and Discussion}

Here we demonstrate the variation of electronic properties of semiconductor SLs due to changes in the lattice strain environments. In particular, we consider electron relaxation in diverse SLs due to scattering with phonons and ionized impurities, and report first-principles calculations of electronic transport properties including the energy-dependent relaxation times. Previous first-principles studies did not include the effect of energy dependent electron relaxation times on electronic transport properties of SLs~\cite{hinsche2012thermoelectric,proshchenko2019optimization,proshchenko2021role,pimachev2021first}. We also discuss how these predictions change when different scattering approximations, such as constant relaxation time and DOS scattering approximations, are used. We investigate two classes of short-period Si/Ge SLs with diverse lattice strain environments: SLs with varied layer compositions grown on identical substrates and SLs with identical compositions but grown on different substrates. We calculate the electronic structure properties of the eight-atom Si/Ge SLs by performing DFT calculations, as implemented in the Quantum Espresso package~\cite{giannozzi2009quantum}. We consider electron relaxation due to scattering with phonons and ionized impurities, and compute both the EPS and IMS rates using a perturbative approach following Fermi's golden rule. We compute electron-phonon matrix elements using the EPW package based on maximally localized Wannier functions~\cite{giustino2007electron,ponce2016epw}, which provide accurate interpolation of the matrix elements from coarse grids to dense grids~\cite{park2014electron} and allows to reduce computational expenses. We perform extensive tests to determine the coarse and fine grid sizes that results in converged electronic properties. We illustrate the physical mechanisms that influence EPS processes and determine the electronic properties of the SLs. Our analysis shows that the energy bands are highly sensitive to the lattice strain environments of the SLs with diverse compositions or SLs grown on different substrates. The layer compositions and substrate induced strains result in non-uniform separations between SL monolayers. The non-uniform layer separations modify the contributions from $\Delta_\perp$ and $\Delta_\parallel$ valleys, and induce energy shift of the SL bands. We illustrate that these energy shifts strongly influence the electron-phonon self-energies. We obtain the EPS rates from the imaginary part of the electron-phonon self-energies. Our analysis establishes a direct relationship between the energy bands and the EPS rates in the SLs with diverse lattice strain environments.

We model the electron-ionized impurity interaction potential with a long-range Coulomb tail that represents a screened ionized impurity. Past first-principles studies demonstrated that incorporation of such long-range Coulomb tail is essential to account for the effect of ionized dopants on electronic transport properties of bulk Si~\cite{qiu2015first}. We show that the EPS processes determine the resistivity of bulk Si at low n$_\text{e}$, while the consideration of IMS processes improves the resistivity predictions at high n$_\text{e}$, matching measured data~\cite{mousty1974relationship}. Our study substantiates the observation that the IMS processes play a significant role in determining electronic resistivities of bulk Si at high carrier concentrations~\cite{fiorentini2016thermoelectric,ponce2016epw}. On the other hand, IMS processes do not significantly affect the diffusive components of the Seebeck coefficients of bulk Si. We further discuss the effect of IMS processes on electron transport properties of Si/Ge SLs, which has not been explored using first-principles approach. We explicitly account for the in-plane and the cross-plane structural anisotropy of the SL configurations, to model the electron-ionized impurity interaction potentials. We obtain anisotropic dielectric tensors from first-principles calculations and use them to compute the electron-ionized impurity perturbation matrix elements. Past studies of IMS rates in bulk Si used isotropic dielectric tensor, and implemented the Brook-Herring's scattering equation~\cite{fiorentini2016thermoelectric,li2012semiconductor}. We illustrate here a modified Brooks-Herring formulation that explicitly considers the effects of anisotropic potentials. However, we note that the \% dielectric anisotropy (defined here as, $|(\epsilon_\perp-\epsilon_\parallel)|/\epsilon_\parallel\times100$), of the chosen SLs stays within 6.5\% and can be considered weak. We report here the values obtained by including the anisotropic dielectric tensor ($\epsilon_\perp$ and $\epsilon_\parallel$), defined in Eq.~\ref{eq:PoissonBoltzmann} with $\epsilon_\perp$ and $\epsilon_\parallel$ shown in Tab.~\ref{tab:dielectric}. Separately, we compute the electronic transport properties using the isotropic BH formalism with an average dielectric constant, $((\epsilon_\perp+2\epsilon_\parallel)/3$). We find that the two sets of predictions do not vary significantly. We do not include the calculations here for brevity. Anisotropic impurity scattering potentials have not been employed in past studies to model scattering rates and resultant SL electronic transport properties. We anticipate that our calculations considering the anisotropic dielectric tensors will encourage future studies of electron-impurity scattering rates in diverse anisotropic systems, going beyond bulk materials.

We demonstrate that the electronic transport coefficients of short-period Si/Ge SLs show significant variations when energy dependent relaxation times are taken into account. We find that the inclusion of the EPS processes negatively impacts the power-factor of the SLs at most $\text{n}_\text{e}$, in comparison with the bulk Si values. Interestingly, when IMS processes are included along with EPS, significant changes are exhibited by the transport properties. We find that Si$_2$Ge$_2$ and Si$_1$Ge$_3$ SLs grown on Si substrate show $~\sim{1.62}$-fold and $~\sim{1.9}$-fold improvement over bulk Si at $\text{n}_\text{e}\sim7.4\times10^{19}\text{cm}^{-3}$ and $\text{n}_\text{e}\sim2.3\times10^{20}\text{cm}^{-3}$, respectively. Their overall peak PFs are $~\sim{1.56}$-fold and $~\sim{1.24}$-fold greater than the bulk Si peak PF. We observe that the PFs can be drastically reduced in SLs with certain lattice strain environments. We illustrate that the electronic and therefore, thermoelectric properties can be tuned by varying the compositions and substrate induced strains of Si/Ge SLs. Although this concept has been previously discussed using analytical and first-principles based CRTA-BTE approaches~\cite{hinsche2011effect,settipalli2020theoretical,proshchenko2021role,proshchenko2019optimization, koga1999carrier}, the energy dependent scattering effects were not considered. Nanostructured Si/Ge SLs are being extensively studied to reduce the thermal conductivity of Si based thermoelectric materials. Additionally, significant reduction of thermal conductivity of Si has been reported at the high doping regime, owing to phonon scattering due to electrons~\cite{liao2015significant}. The power-factor improvements at high $\text{n}_\text{e}$ predicted by our study, could provide promising avenues for thermoelectric applications, when combined with low thermal conductivities. We hope that our results will encourage experimental studies to validate this promising physical phenomenon, illustrated by taking into account the effect of different energy dependent scattering processes. Our results establish the role of the strain-modulated DOS, electron group velocities and energy-dependent electron scattering rates on the electronic transport properties. The physical insights demonstrated in this analysis not only are valuable to predict band movement and electron relaxation in complex heterostructures, but essential to develop strain engineering approaches to optimize electronic properties. For example, understanding of the interactions of electrons with real and quasiparticles is crucial to optimize the electronic properties of thermoelectric and other technology enabling materials.

\section*{Conflicts of interest}
There are no conflicts to declare.

\section*{Acknowledgements}
We are indebted to Samuel Ponc{\'e} for useful discussions and assistance with implementing EPW code for electron-phonon scattering rate calculations. We gratefully acknowledge funding from the Defense Advanced Research Projects Agency (Defense Sciences Office) [Agreement No.: HR0011-16-2-0043]. This work used the Extreme Science and Engineering Discovery Environment (XSEDE), which is supported by National Science Foundation grant number ACI-1548562. We are grateful for the computing resources provided by the Department of Defense (DoD) High Performance Computing Modernization Program (HPCMP) at the Air Force Research Laboratory DoD Supercomputing Research Center (AFRL-DSRC). We acknowledge the computing resources provided the RMACC Summit supercomputer, which is supported by the National Science Foundation (awards ACI-1532235 and ACI-1532236), the University of Colorado Boulder, and Colorado State University. The Summit supercomputer is a joint effort of the University of Colorado Boulder and Colorado State University.

\bibliography{EPhLiterature} 

\end{document}